\documentclass[namedreferences]{solarphysics}

\usepackage[hyperref,optionalrh]{spr-sola-addons} 
\usepackage{graphicx}        
\usepackage{color}           

\renewcommand{\vec}[1]{{\mathbfit #1}}

\chardef\us=`\_

\begin{document}

\begin{article}
\begin{opening}

\title{Meridional Motions and Reynolds Stress Determined by Using Kanzelh\"ohe Drawings and
White Light Solar Images from 1964 to 2016}

\author[addressref={aff1},corref,email={rdomagoj@geof.hr}]{\inits{D.}\fnm{Domagoj}~\lnm{Ru\v{z}djak}}\sep
\author[addressref={aff1}]{\inits{D.}\fnm{Davor}~\lnm{Sudar}}\sep
\author[addressref={aff1}]{\inits{R.}\fnm{Roman}~\lnm{Braj\v{s}a}}\sep
\author[addressref={aff1}]{\inits{I.}\fnm{Ivica}~\lnm{Skoki\'c}}\sep
\author[addressref={aff2}]{\inits{I.}\fnm{Ivana}~\lnm{Poljan\v{c}i\'c Beljan}}\sep
\author[addressref={aff2}]{\inits{R.}\fnm{Rajka}~\lnm{Jurdana-\v Sepi\'c}}\sep
\author[addressref={aff3}]{\inits{A.}\fnm{Arnold}~\lnm{Hanslmeier}}\sep
\author[addressref={aff3,aff4}]{\inits{A.}\fnm{Astrid}~\lnm{Veronig}}\sep
\author[addressref={aff4}]{\inits{W.}\fnm{Werner}~\lnm{P\"otzi}}
\address[id=aff1]{Hvar Observatory, Faculty of Geodesy, University of Zagreb, Ka\v{c}i\'ceva 26, 10000 Zagreb, Croatia}
\address[id=aff2]{Department of Physics, University of Rijeka, Radmile Matej\v{c}i\'c 2, 51000 Rijeka, Croatia}
\address[id=aff3]{IGAM - Institute of Physics, University of Graz, Universit\"atplatz 5, 8010 Graz, Austria}
\address[id=aff4]{Kanzelh\"ohe Observatory for Solar and Environmental Research, Kanzelh\"ohe 19, 
9521 Treffen am Ossiacher See, Austria}

\runningauthor{D. Ru\v{z}djak {\it et al.}}
\runningtitle{Meridional Motions and Reynolds Stress from Kanzelh\"ohe Data}

\begin{abstract}
Sunspot position data obtained from Kanzelh\"ohe Observatory for Solar and Environmental Research
(KSO) sunspot drawings and
white light images in the period 1964 to 2016 were used to calculate
the rotational and meridional velocities of the solar plasma.
Velocities were calculated from daily shifts of sunspot groups and
an iterative process of calculation of the differential rotation profiles
was used to discard outliers. We found a differential rotation profile
and meridional motions in agreement with previous studies using sunspots as 
tracers and conclude that the quality of the KSO data is appropriate for 
analysis of solar velocity patterns. By analysing the correlation and
covariance of meridional velocities and rotation rate residuals
we found that the angular momentum is transported towards the solar 
equator. The magnitude and latitudinal dependence of the horizontal
component of the Reynolds stress tensor calculated is sufficient 
to maintain the observed solar differential rotation profile.
Therefore, our results confirm that the Reynolds stress is the 
dominant mechanism responsible for transport of angular momentum 
towards the solar equator. 
\end{abstract}
\end{opening}


\section{Introduction}

Precise determination of solar large scale velocity patterns can provide information 
about the transport of angular momentum in the solar convective zone and provide important 
observational constraints for the solar dynamo models.
One possibility to determine the solar velocity field is by observing the motions of 
structures which can be observed at the surface of the Sun. Most often sunspots
and sunspot groups were used as tracers \citep[$e.g.$][among many others]{howard1984,
balthasar1986,howard1991,lustig1994,pulkkinen1998a,woehl2001,zuccarello2003,sudar2014,
sivaraman2010,mandal2017,sudar2017}. Besides tracing sunspots, other methods have been used 
for assessment of solar large scale flows, for instance: Doppler measurements 
\citep[$e.g.$][]{hathaway1996} and tracing coronal bright points (CBP) 
\citep[$e.g.$][]{sudar2016}. In recent years the observations of solar velocity field were 
revolutionized by helioseismology \citep{hanasoge2015}. 
While all mentioned methods give very similar results for solar rotation the obtained
results for meridional flows are controversial, as described in
\citet{hathaway1996} and \citet{sudar2017}. 
The Doppler measurements as well as observation and analysis of global oscillations 
reveal that there is a poleward meridional circulation in the near surface layers of the 
Sun in both hemispheres. This is in agreement with the result of most theoretical models 
which predict unicellular meridional circulation directed poleward at the top and
equatorward at the bottom of the convection zone \citep{brun2009}.
Observations utilizing tracers show more complicated pattern of meridional flows.
All kinds of meridional flow directions were found (poleward, equatorward, towards and away from the 
center of activity). However the results for meridional circulation using tracers are influenced 
by several effects. First, the active regions locally modify the amplitude and direction
of meridional circulation \citep{haber2004,svanda2008}. Next, the movement of the (magnetic) 
tracers do not represent the movement of the solar surface plasma, but the movement of the 
layer where the observed features are anchored, which might change with time \citep{ruzdjak2004},
and finally, the solar meridional circulation might be variable as pointed out by \citet{hathaway1996}.

Differential rotation of the Sun can be explained as rotationally influenced turbulence in the 
convective zone. The turbulence leads to the formation of large-scale turbulent fluxes 
\citep{Rudiger2004}. The angular momentum fluxes are proportional to the velocity correlation 
tensor and are given by:
\begin{equation}
q_{ij}=\overline{v^\prime_i v^\prime_j}
\end{equation}
where $q_{ij}$ is Reynolds stress tensor, $\vec v$ is velocity, the overbar denotes azimuthal averaging, 
and primes denote variations about the averages. The latitudinal flux of the angular momentum
is described by the horizontal component of the Reynolds stress tensor $q_{\lambda b}$, 
which can be calculated as the covariance of the meridional motion and the rotation velocity residuals. 
The rotation and meridional circulation of tracers can easily be determined separately.
Therefore, contrary to meridional flow analysis, tracers are suitable tool for analysing 
the latitudinal flux of the angular momentum, {\it i.e.} the turbulent Reynolds stress as the main driver of differential rotation.

Kanzelh\"ohe Observatory for Solar and Environmental Research (KSO) was 
founded during WW II as one station within a network of observatories 
for  observing ``solar eruptions" (flares) which were interfering with
radio communications. 
Nowadays KSO is affiliated with the University of Graz and performs regular 
high-cadence full-disk observations of the 
Sun in the H$\alpha$, the Ca {\sc ii} K spectral lines, and in white
light with a coverage of about 300 observing days {\it per} year \citep{veronig2016}.

KSO white light images and sunspot drawings have been used by different authors for measuring
the photospheric velocity fields.  The data before 1985 were used, 
{\it e.g.} by \citet{lustig1983}, \citet{hanslmeier1986}, \citet{lustig1987},
\citet{balthasar1988} and \citet{lustig1991}. 
\citet{poljancic2010} and \citet{poljancic2011} compared the  GPR, USAF/NOAA, 
DPD and KSO sunspot databases and found that DPD and KSO data are, in some respect, more 
accurate than the USAF/NOAA data. Consequently, the venture of determination
of the heliographic positions from the sunspot drawings and full disc white 
light CCD images was undertaken. The procedure and results for solar rotation 
in the period 1964-2016 are presented in \citet{poljancic2017}.
Here we present the analysis of meridional motions and Reynolds stress determined 
from KSO data in the same period 1964-2016.


\section{Data and Analysis}

The drawings of the whole Sun are made at KSO using a refractor telescope ($d/f$=110/1650 mm).
Additionally, from 1989 onwards the white light photographs of the whole Sun are made 
with a refractor telescope ($d/f$=130/1950 mm), where the photographic camera was replaced with a 
CCD camera in July 2007. The positions of the sunspot groups were measured by two methods:
interactive and automatic.
The interactive procedure was applied for data from 1964 to 2008, where the ``Sungrabber" software 
package \citep{sungrabber} was used by two independent observers to measure the positions of group centers
on the sunspot drawings made at KSO.
For the automatic procedure, morphological image processing based on the ``STARA" algorithm \citep{watson2011} 
was used for the determination of the positions of sunspot groups. The automatic method was applied 
to the data observed with the digital cameras first of which was installed in July 2007.  
Since the Solar Cycle 23 ended in 2008 to have an homogeneous dataset within the solar cycle only 
the data during 2009-2016 were obtained by the automatic method. A detailed description of both 
methods and the availability of the data is given in \citet{poljancic2017}.

To check how the two methods compare with each other the 
drawings of the whole Sun made during 2014 (solar maximum) were measured using Sungrabber. 
Descriptive statistics of meridional motions and rotation 
rate residuals calculated using both methods are presented in
Table \ref{statistic}.

  \begin{table}[h]
      \caption{The 
measures of central tendency and dispersion for meridional motions and rotation 
rate residuals obtained by interactive and automatic methods. Prior to calculation 
 the 
outliers were discarded. Stdev stand for standard deviation, IQR for interquartile range,
Skew for skewness and Kurt for kurtosis}
         \label{statistic}   
         \begin{tabular}{lrcrrrrrr}
            \hline
   Quantity & Method  & N & Mean & Median & Stdev & IQR & Skew & Kurt   \\

    $v_\mathrm{mer}$ (m\,s$^{-1}$) & interactive & 961 &2 &--1& 74 & 83& 0.08& 1.7 \\
    $v_{\rm mer}$ (m\,s$^{-1}$) & automatic   & 792 &--1&--1&73&57&--0.12&4.2 \\
$\Delta v_{\rm rot}$ (m\,s$^{-1}$)& interactive & 961 &76&77&167&207&--0.12&0.91 \\
$\Delta v_{\rm rot}$ (m\,s$^{-1}$)& automatic   & 792 &5&-6&182&144&0.04&1.4 \\
                       \hline
         \end{tabular}
   \end{table}

A data set of 45914 times and positions of sunspot groups during the period from January 1964 to April 2016
were used to calculate meridional and rotational speeds. For the sunspot groups for which the 
Central Meridian Distance (CMD) was less than 58$^\circ$, which corresponds to about 0.85
of the projected solar radius \citep{balthasar1986}, rotation speeds were calculated by division 
of CMD differences by elapsed time and meridional motions were calculated by division of
latitude differences by elapsed time. This resulted in 33817 rotation and
meridional velocity values. The obtained synodic rotation velocities were transformed 
to sidereal ones by the procedure described in \citet{skokic2014}. 
Finally, to account for errors of 
misclassification and other errors, an iterative fitting method was used, similar to the one
used in \citet{sudar2016} and \citet{sudar2017}. 

Rotation rate residuals were calculated by 
subtracting the individual rotation velocities from the average rotation profile:
\begin{equation} 
\omega(b)=A+B\sin^2b,
\label{rotprofile}
\end{equation}
where $A$ and $B$ are differential rotation parameters in [$^\circ$\,day$^{-1}$] and $b$ is the
heliographic latitude in [$^\circ$].
Robust statistics of the rotation rate residuals and meridional velocity was used, 
and values lying outside 3.5 interquartile ranges from the median were considered 
as outliers and discarded. Since the removed outliers were contributing to the mean 
rotation profile derived, the process is iteratively repeated until no outliers are 
present in the data. The procedure converges very fast and after 4 iterations
no outliers were present. Data whose absolute values of rotation rate 
residual and meridional velocity were larger than 4.2$^\circ$\,day$^{-1}$ and 2.3$^\circ$\,day$^{-1}$,
respectively, were discarded. After applying all these reduction steps 32616 data points
are left for further analyses. 

\begin{figure}
\centering
   \includegraphics[width=0.8\textwidth]{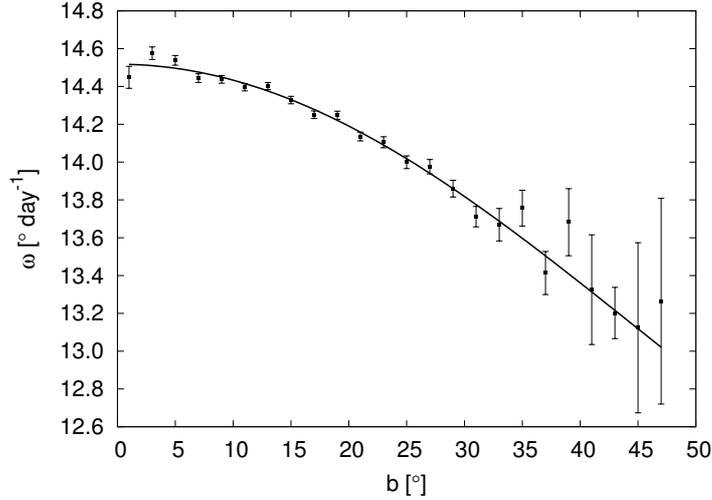}
      \caption{Differential rotation profile obtained from KSO data from 1964 to 2016. 
Points with error 
bars are averaged 2$^\circ$ latitude bin values and the best fit profile (Equation \ref{rotprofile}) is 
shown with the solid line.}
         \label{rprofile}
   \end{figure}

In Figure \ref{rprofile} the 
obtained differential rotation profile is presented, the best fit differential 
rotation parameters are: $A=14.5177\pm 0.0096^\circ$\,day$^{-1}$ and
$B=-2.800\pm 0.088^\circ$\,day$^{-1}$. The  averaged 2$^\circ$ latitude bin
values of $\omega(b)$ are also shown. The errors for bins at higher 
latitudes are quite large due to the small number of sunspots present at these latitudes.

When analysing latitudinal dependencies the latitude of the first measurement was assigned 
to each rotational and meridional velocity \citep{olemskoy2005} as was done in 
\citet{sudar2014,sudar2015,sudar2016,sudar2017} to avoid false meridional flows. 
The rotation rate residuals and meridional 
velocities were transformed from angular values to linear ones. Taking $R_\odot=6.96\times 10^8$m,
the conversion factors are 140.6 and 140.6\,$\cos(b)$ m\,$^{-1}$\,day($^\circ$)$^{-1}$ 
for meridional velocities and rotation velocity residuals, respectively, where the latitude
of the first measurement was taken into account. In addition the meridional speeds are 
transformed so that negative value of meridional speed represents motion toward the 
equator for both solar hemispheres. This is achieved by changing the sign of meridional
velocities for the southern solar hemisphere, where negative values of latitude are assigned.

\section{Results}
\subsection{Latitudinal Dependence of Meridional Motions and Rotation Velocity Residuals}

   \begin{figure}
\centering
   \includegraphics[height=0.33\textwidth,angle=-90]{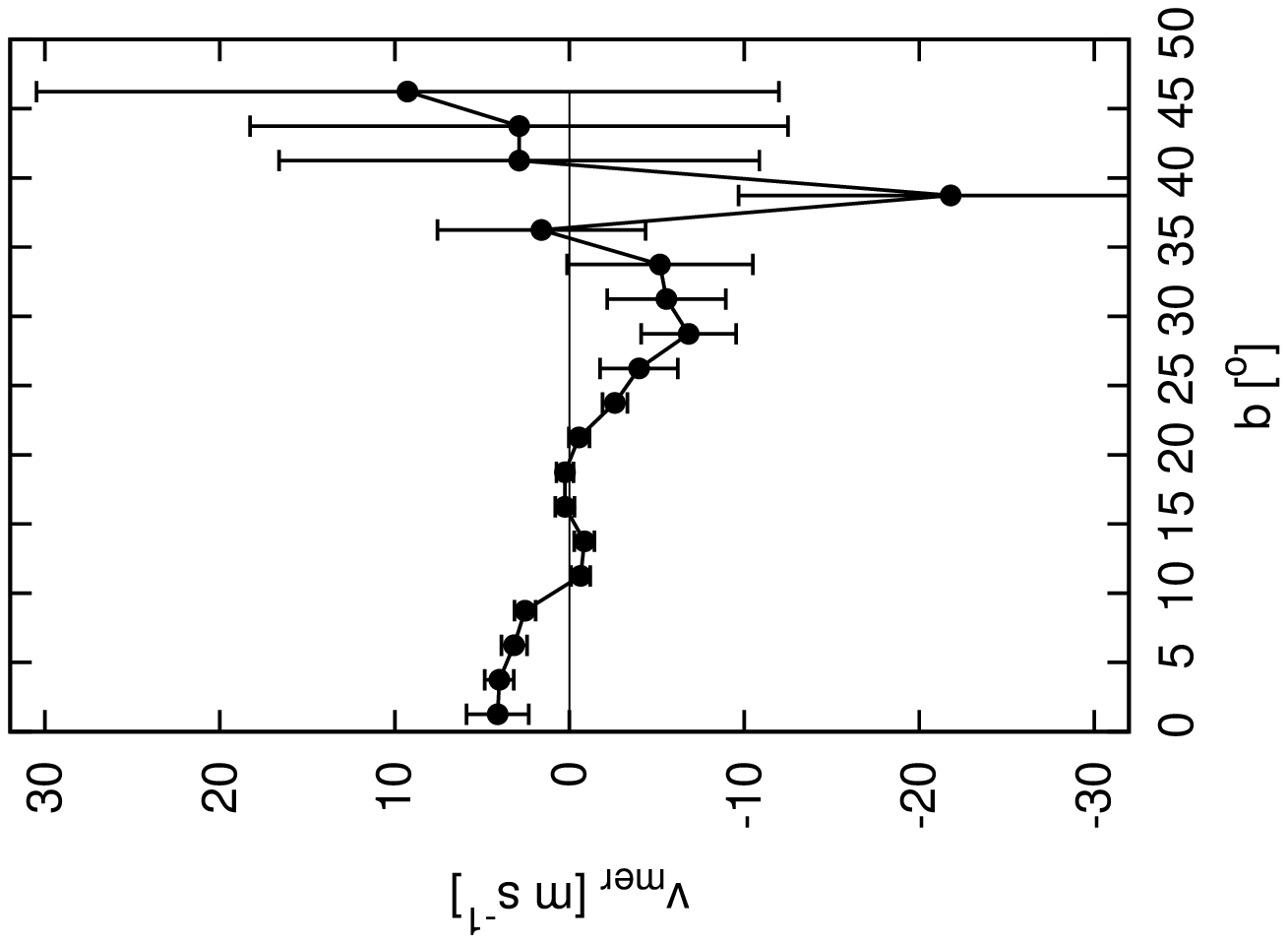}
   \includegraphics[height=0.33\textwidth,angle=-90]{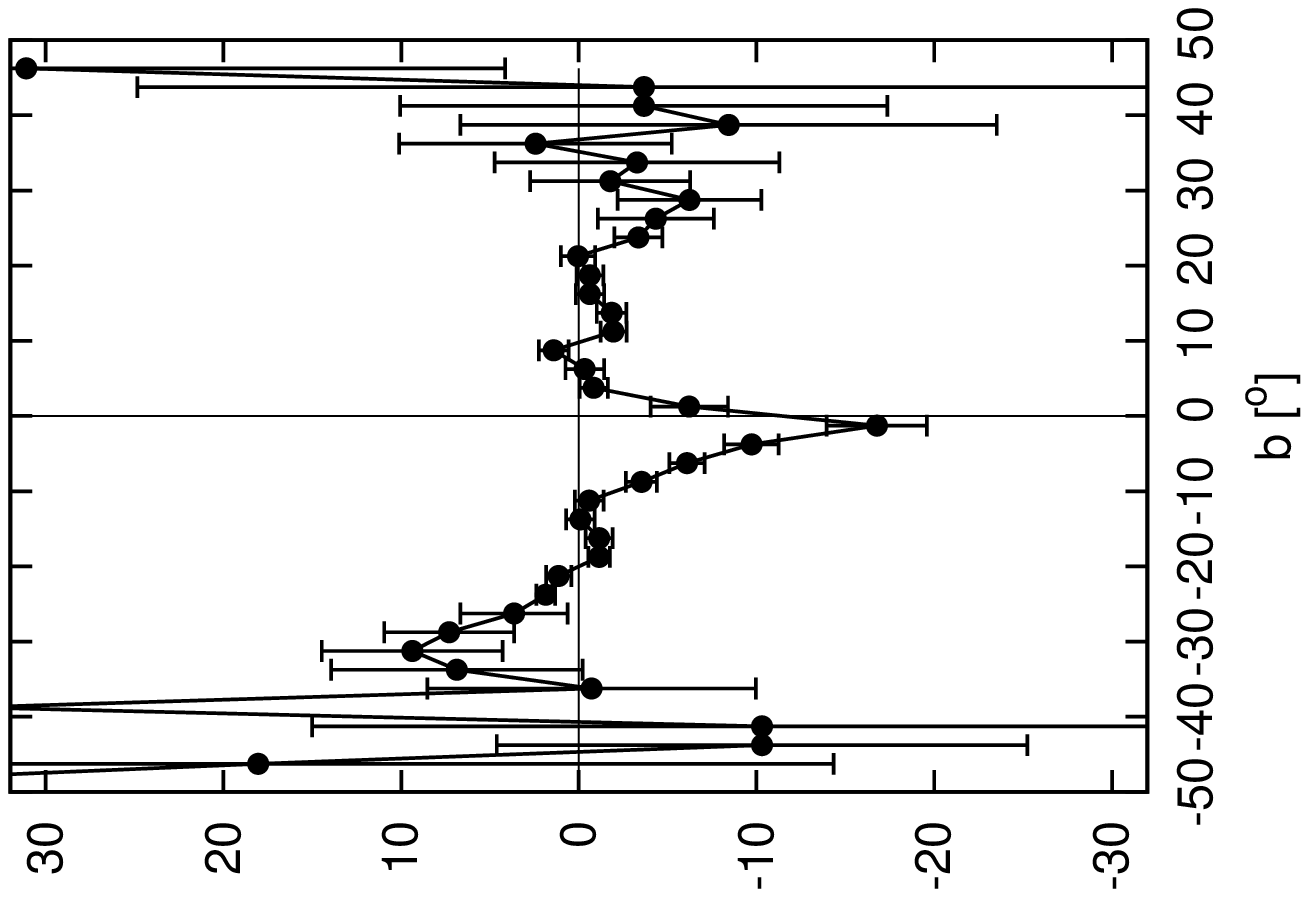}
      \caption{Meridional motions as a function of latitude. Data are averaged 
over 2.5$^\circ$ in latitude. In the left panel both solar hemispheres are shown together
and positive values indicate motion towards the poles. In the right panel the northern and southern
hemisphere are shown separately and positive values indicate motion towards north. }
         \label{mer}
   \end{figure}

  \begin{figure}
\centering
   \includegraphics[width=0.8\textwidth]{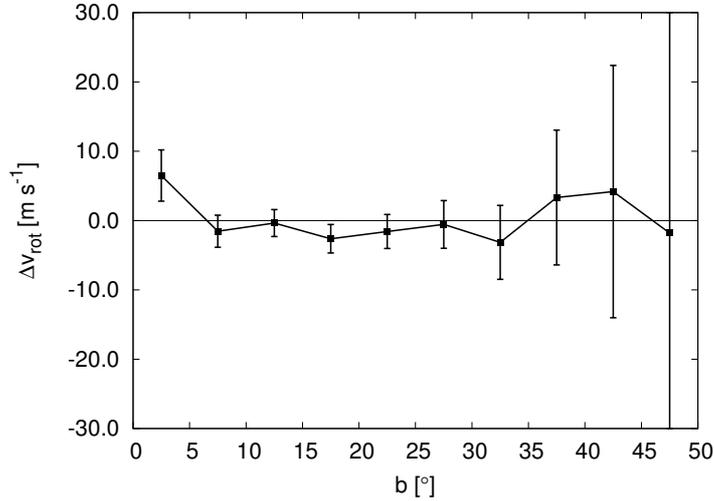}
      \caption{Rotation velocity residuals as function of latitude. Data were averaged over 
5$^\circ$  in latitude. Positive values denote rotation faster than average and negative values
rotation slower than average. Both solar hemispheres have been treated together.}
         \label{torsional}
   \end{figure}

The dependence of meridional motions obtained from the KSO data on latitude is illustrated in 
Figure \ref{mer}. The data were averaged over 2.5$^\circ$ in latitude and mean values with
error bars are given for each latitude stripe. It can be seen that at low latitudes ($\leq10^\circ$)
meridional motions are toward the poles. The latitude stripes 20$^\circ$--30$^\circ$ shows 
motions toward the solar equator, while the rest of the values are not significantly different 
from zero. 
In the right panel of the figure each solar hemisphere is shown separately. 
Here the meridional velocities are not transformed and  positive values of the meridional 
velocity denotes motion towards north.
The result for the southern hemisphere shows motions toward the pole at low latitudes and
changes to flow towards north (equator) at higher latitudes. This behavior is reminiscent 
of the one found by  \citet{sudar2014} analysing GPR, USAF/NOAA data and by \citet{sudar2017} 
studying DPD data.
The values for the northern solar hemisphere are not significantly different from zero in all
latitude stripes.
The most statistically significant values are for stripes 0$^\circ$--2.5$^\circ$ and 
22.5$^\circ$--25$^\circ$ both showing motions toward south. The observed meridional motions would 
be consistent with the equatorward motions on the northern solar hemisphere. 
Such motions were observed on both solar hemispheres by \citet{sivaraman2010} 
analysing Kodaikanal and Mt. Wilson data.

Figure \ref{torsional} shows the dependence of rotation residual velocities on latitude. 
The data were averaged over 5$^\circ$ in latitude and average values 
with error bars are presented for each latitude stripe. None of the rotation rate 
residual values is significantly different from zero.

  \begin{table}[h]
      \caption{Description of data subsets with cycle and phase boundaries. Slope and
both intercept values are the result of meridional velocities latitude dependence linear fit.
The solar cycle boundaries are taken from \citet{brajsa2009}.}
         \label{subset}   
         \begin{tabular}{lcccc}
            \hline
     Description & Boundaries  & Slope & Intercept\ Y & Intercept\ X   \\
                       \hline
Solar Cycle 20 &21.10.1964--20.04.1976 & $-0.42\pm0.19$ & $8.0\pm2.8$ & $19.0\pm10.0$ \\
Solar Cycle 21 &21.04.1976--15.09.1986 & $-0.37\pm0.13$ & $4.7\pm2.2$ & $12.7\pm\ 7.4$ \\
Solar Cycle 22 &16.09.1986--25.05.1996 & $-0.20\pm0.11$ & $2.1\pm1.9$ & $10.5\pm11.1$ \\
Solar Cycle 23 &26.05.1996--30.06.2008 & $-0.43\pm0.11$ & $6.7\pm1.9$ & $15.5\pm\ 6.0$ \\
\hline
Minimum &02.01.1964--25.05.1967 &  & &   \\
from 2y before &21.04.1974--30.06.1978 & & &  \\
minimum till &16.09.1984--31.12.1987 &$-0.19\pm0.14$ & $3.4\pm2.5$ & $17.9\pm18.6$ \\
1.5y before &26.05.1994--20.10.1998 &  & &   \\
maximum     &01.07.2006--25.11.2012 &  & &   \\
\hline
Pre maximum &26.05.1967--25.11.1968 & & &  \\ 
from 1.5 y &01.07.1978--31.12.1979 & & &  \\
prior to &01.01.1988--30.06.1989 & $-0.45\pm0.13$ & $7.8\pm2.7$ & $17.3\pm7.8$ \\
maximum till &21.10.1998--20.04.2000 &  & &  \\
maximum & 26.11.2012--25.05.2014 & & &  \\
\hline
Past maximum &26.11.1968--25.05.1970 & & &  \\
from maximum &01.01.1980--30.06.1981 & & &  \\
till 1.5 y &01.07.1989--31.12.1990 & $-0.33\pm0.12$ & $4.4\pm2.1$ & $13.3\pm8.0$ \\
after the &21.04.2000--20.10.2001 & & &  \\
maximum   &26.05.2014--20.04.2016 & & &  \\
\hline
Declining phase &26.05.1970--20.04.1974 & & &  \\ 
Fom 1.5y after &01.07.1981--15.09.1984 &$-0.45\pm0.12$ & $5.1\pm1.6$ & $11.3\pm4.7$ \\
max. till 2y&01.01.1991--25.05.1994 & & &  \\
 before minimum &21.10.2001--30.06.2006 & & &  \\
                       \hline
         \end{tabular}
   \end{table}

   \begin{figure}
\centering
   \includegraphics[height=0.24\textwidth,angle=-90]{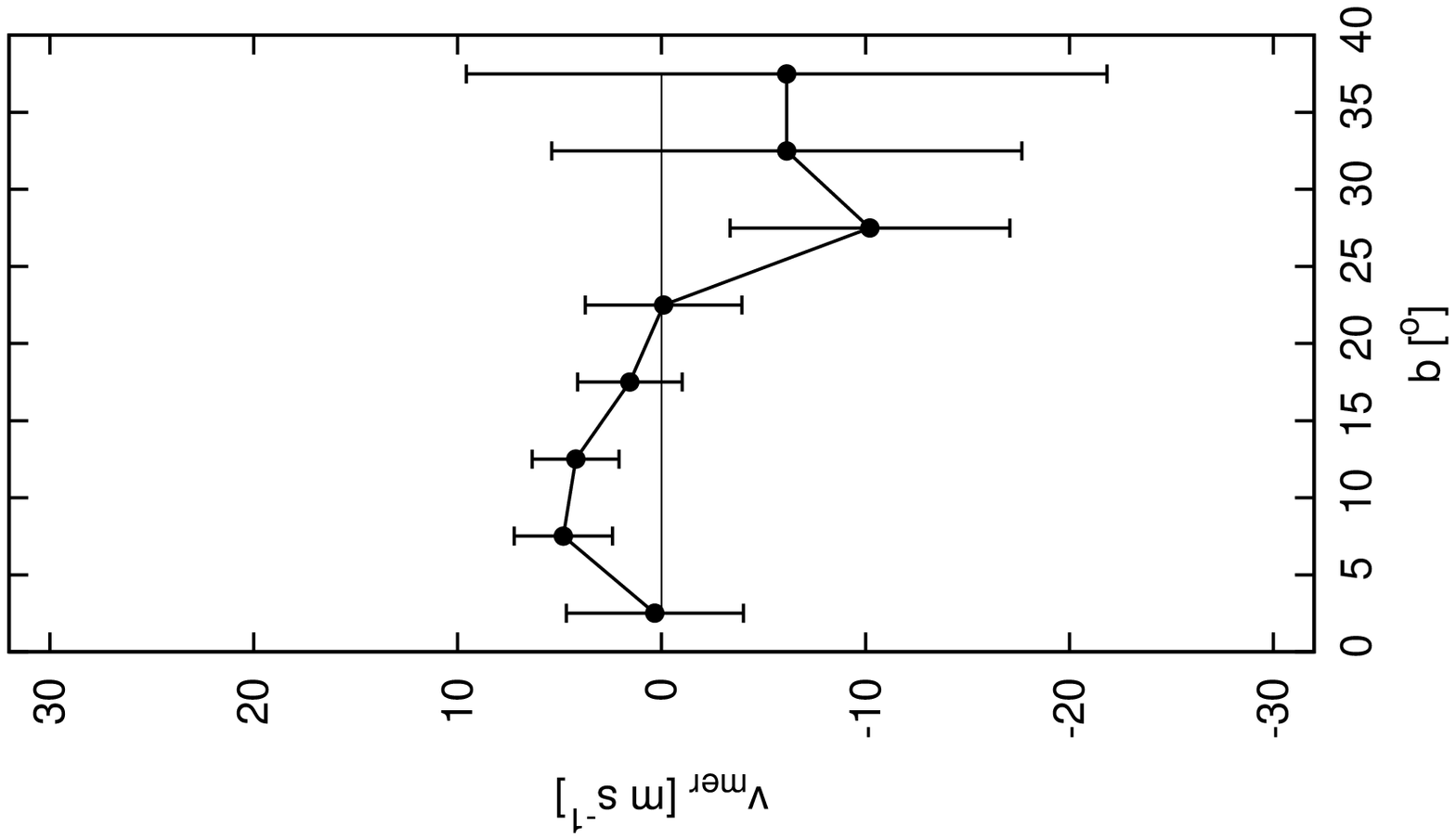}
   \includegraphics[height=0.24\textwidth,angle=-90]{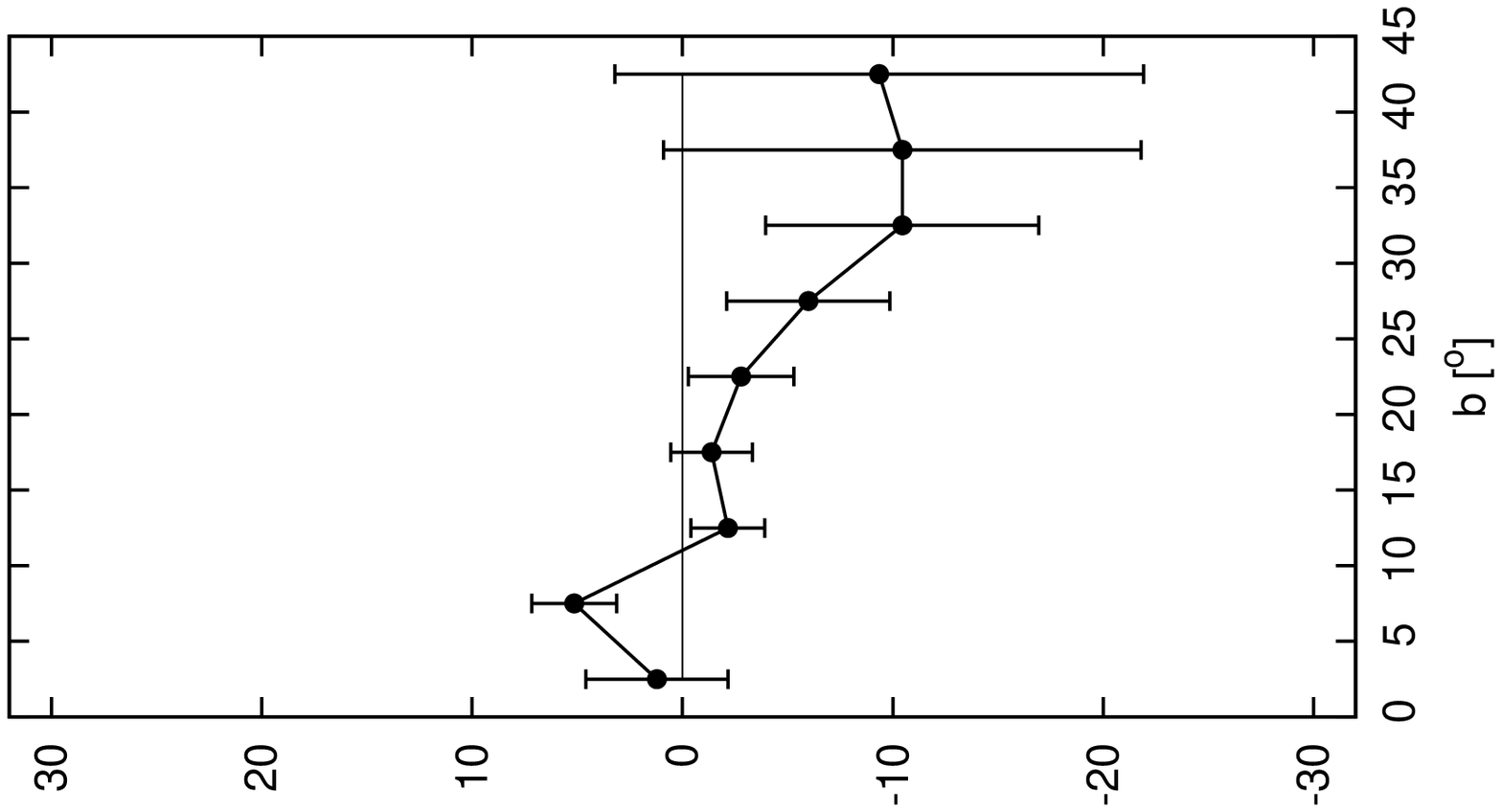}
   \includegraphics[height=0.24\textwidth,angle=-90]{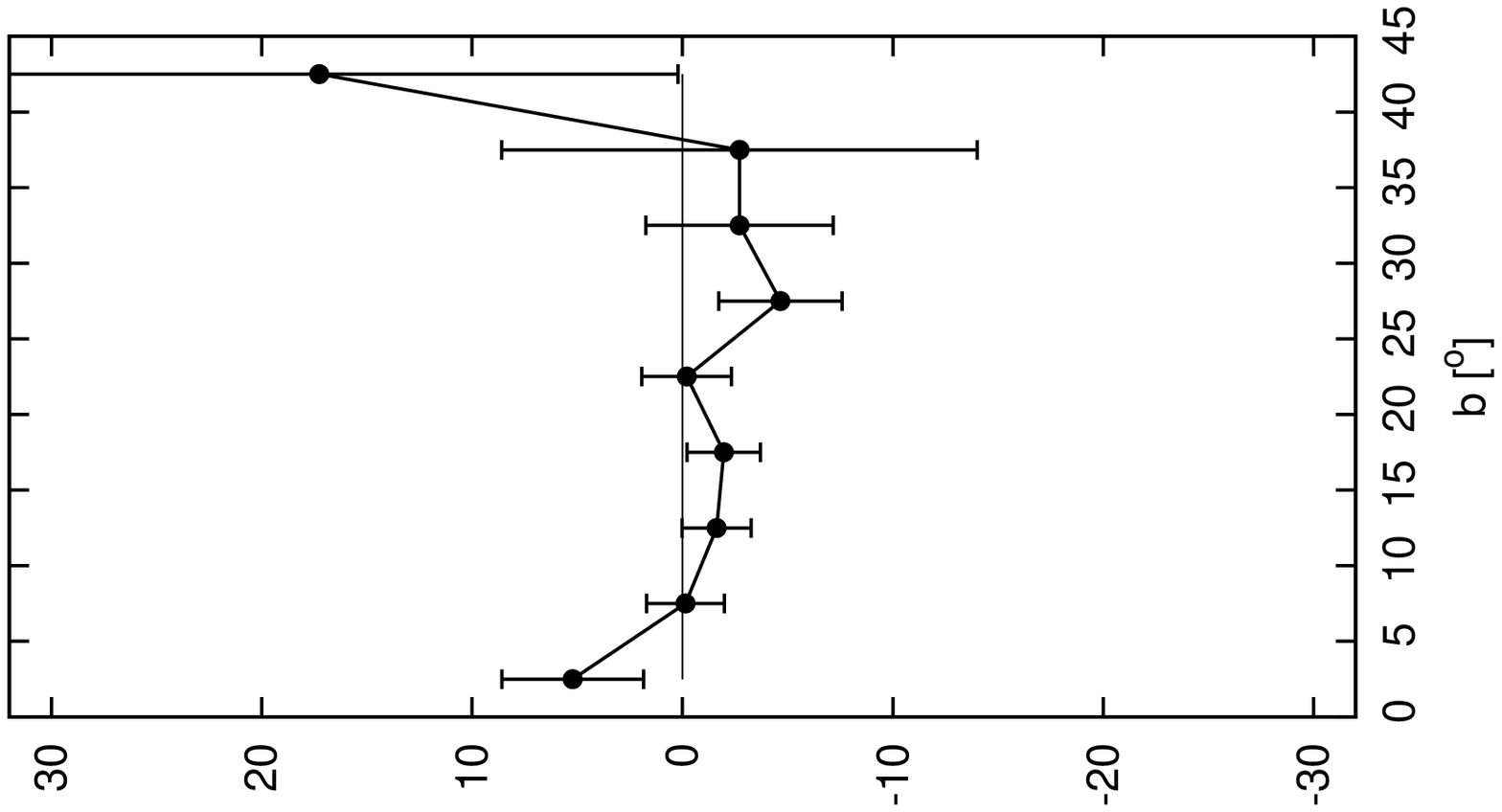}
   \includegraphics[height=0.24\textwidth,angle=-90]{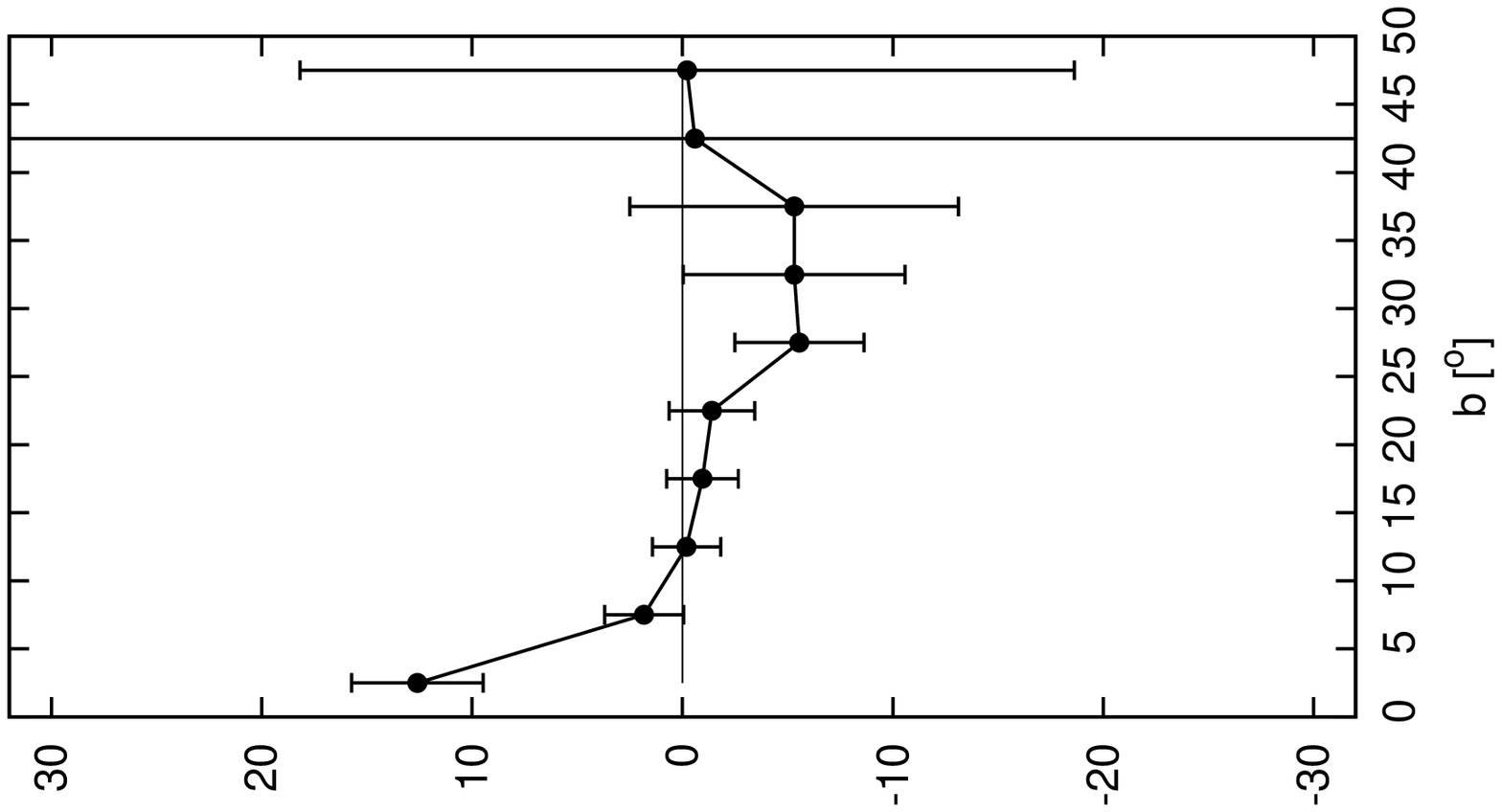}
   \includegraphics[height=0.24\textwidth,angle=-90]{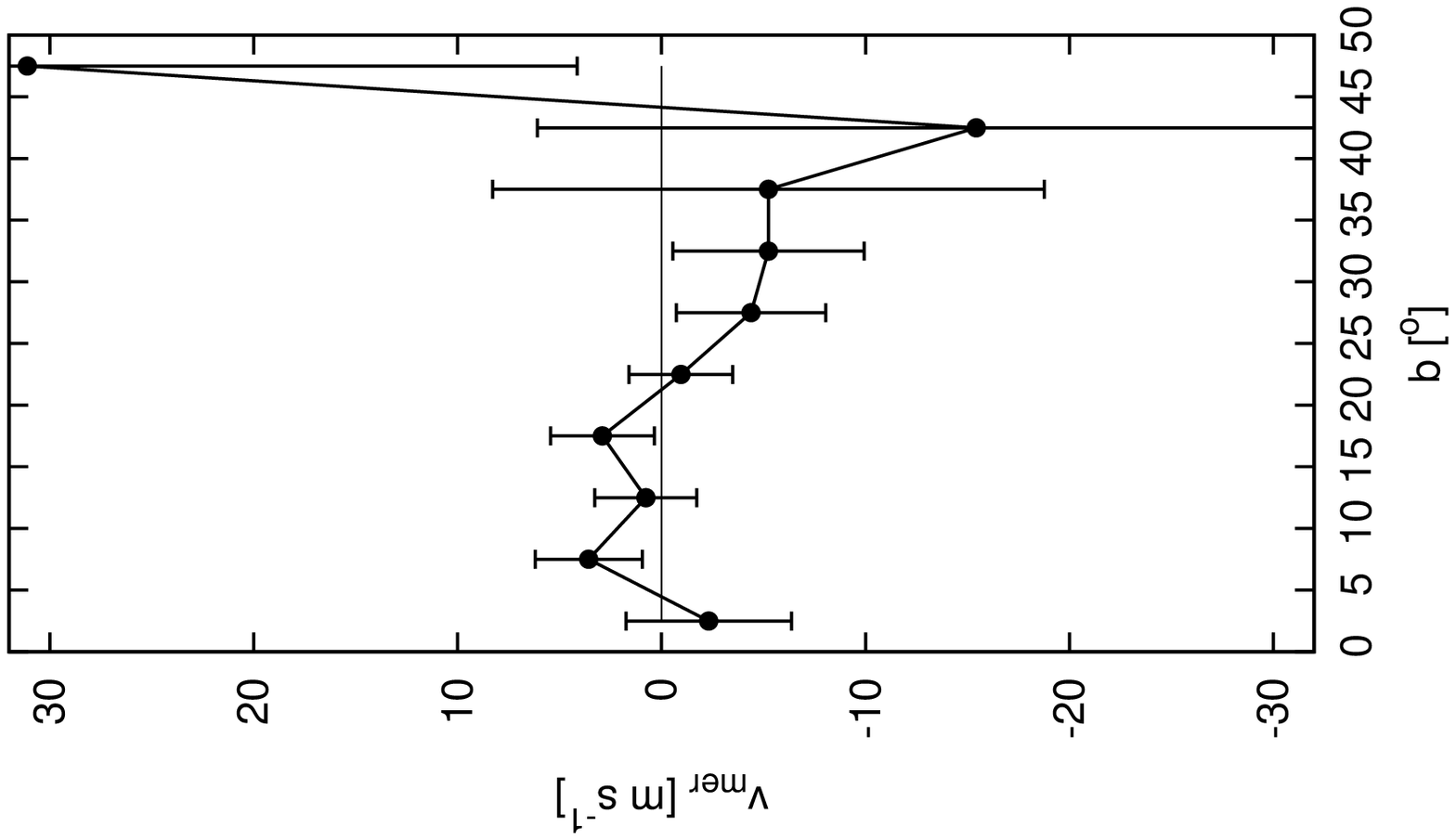}
   \includegraphics[height=0.24\textwidth,angle=-90]{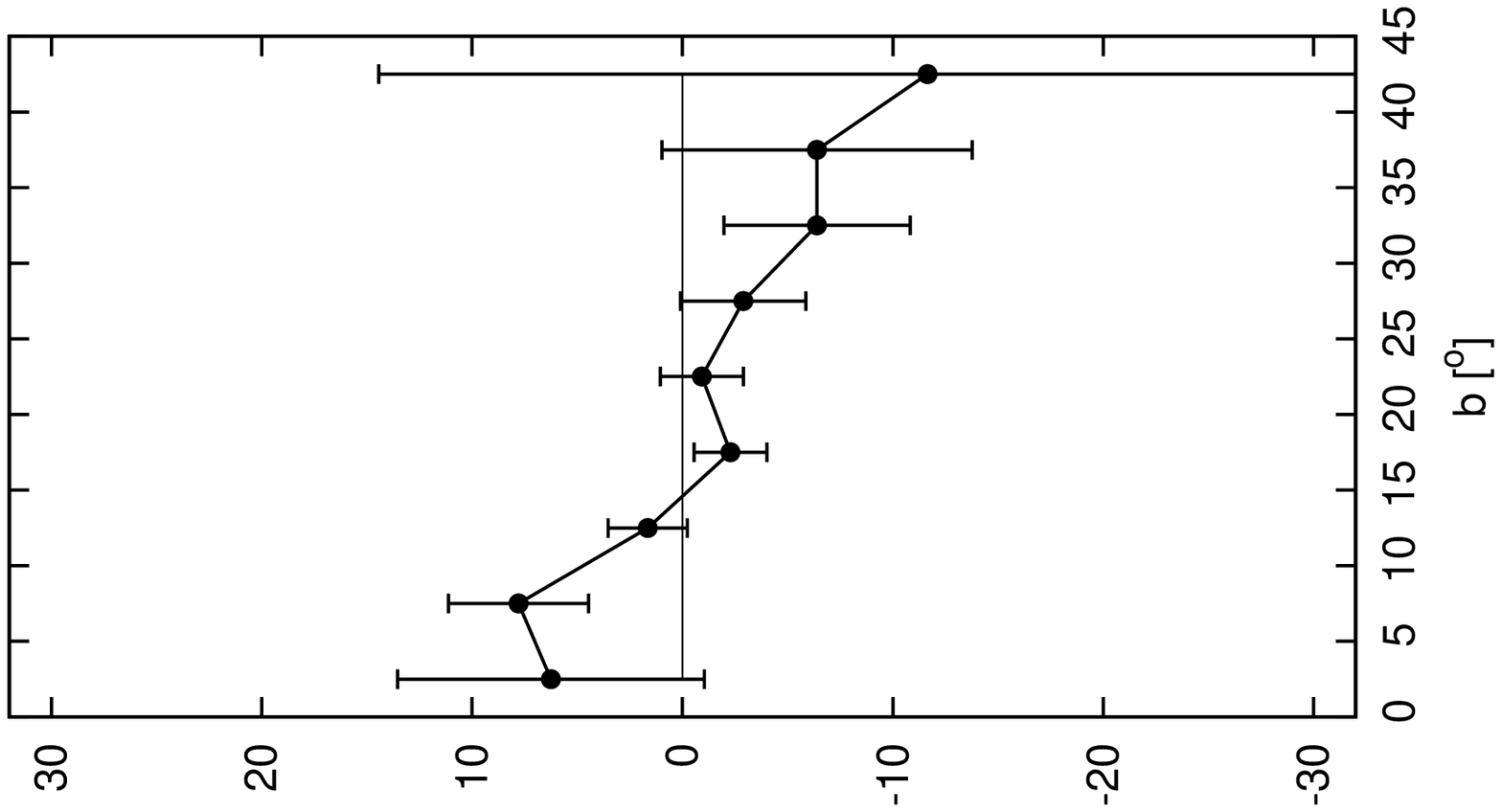}
   \includegraphics[height=0.24\textwidth,angle=-90]{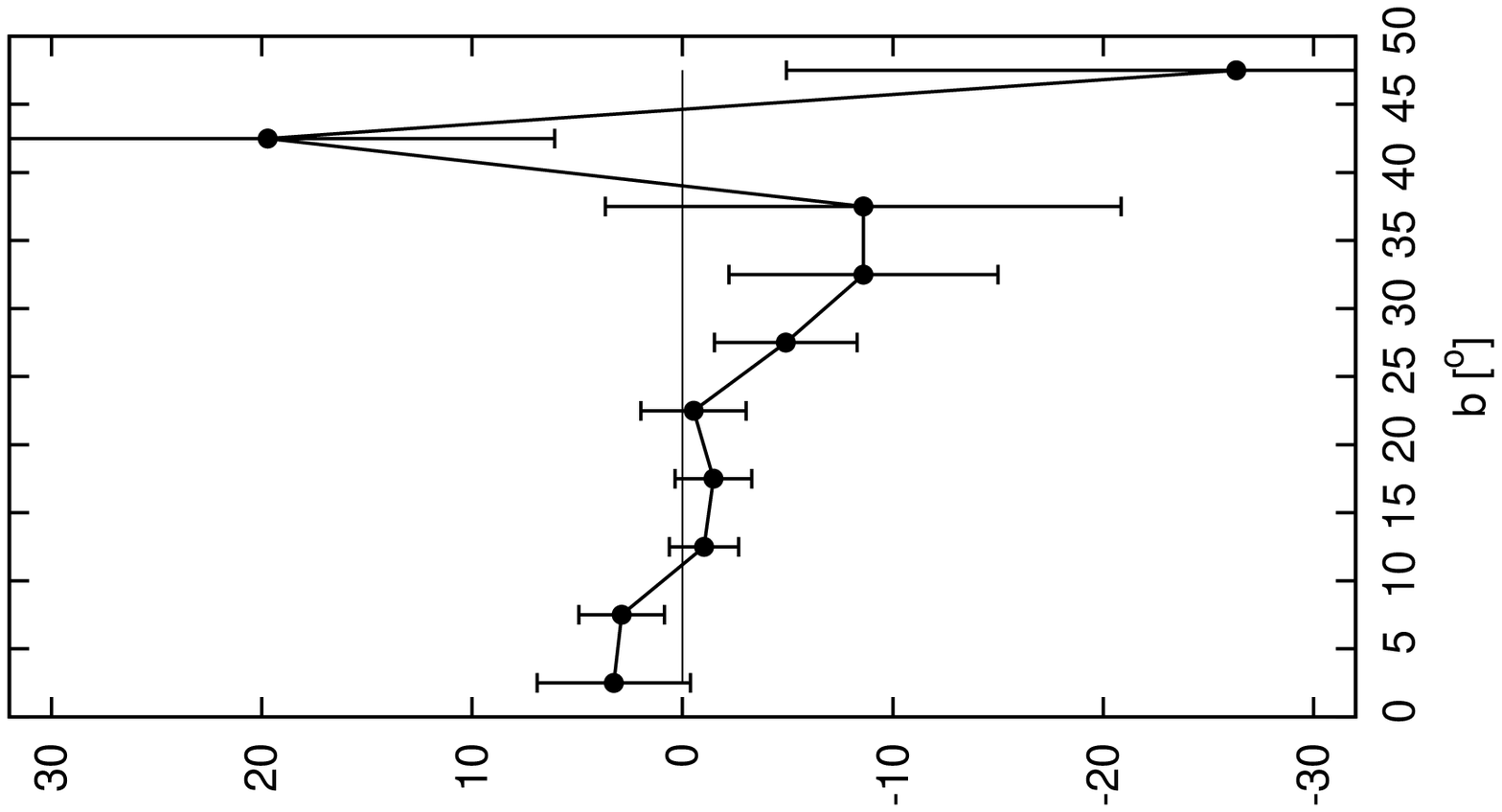}
   \includegraphics[height=0.24\textwidth,angle=-90]{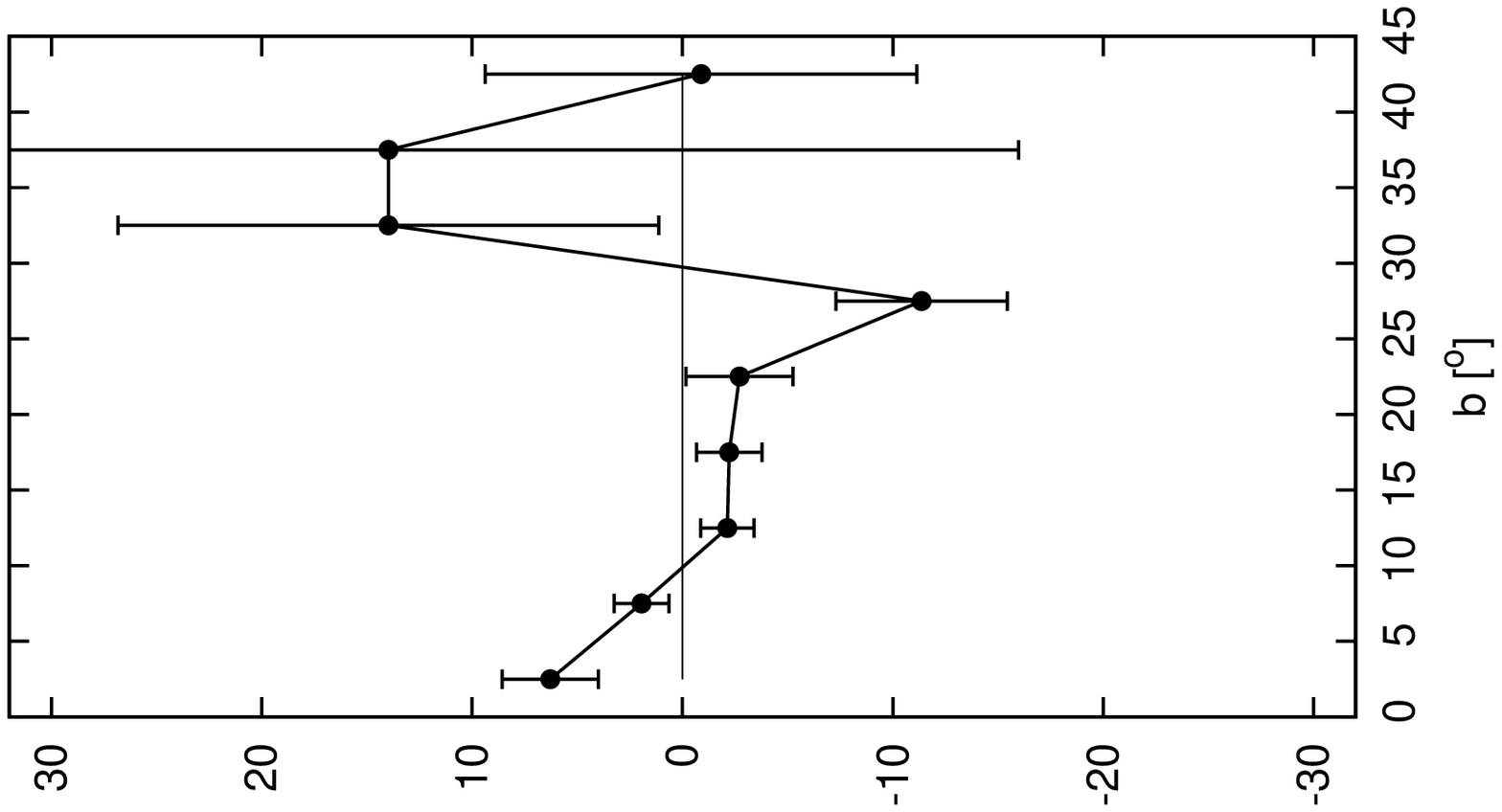}
      \caption{Meridional motions as a function of latitude. In the upper row profile for each 
cycle from Solar Cycle 20 to 23 is shown separately (from left to right). 
In lower row four different phases of the cycle are presented (minimum, pre-maximum, past-maximum 
and declining phase, from left to right). 
Velocities belonging to corresponding phase from all cycles were averaged. 
Data are averaged over 5$^\circ$ in latitude and both solar hemispheres 
are shown together to have sufficient number of data in each latitude bin. 
Positive values indicate motion towards the poles.}
         \label{cycle}
   \end{figure}

To examine the changes of meridional circulation with time and the phase of the solar cycle
the dataset was divided into four subsets containing individual cycles from Solar Cycle 20 
to Solar Cycle 23 and four subsets corresponding to different phases of the cycle. The description 
of the data subsets is given in Table \ref{subset}. To have sufficient number of data in each latitude
stripe both solar hemispheres were treated together and data were averaged over 5$^\circ$ in latitude.
The latitudinal dependence of meridional motions for individual solar cycles is presented in 
upper row of Figure \ref{cycle} and the changes within the cycle are shown in the lower row.
All meridional velocity profiles showing the motions toward pole at low latitudes and 
motions toward equator at higher latitudes. The exception is the profile observed during solar 
cycle minimum.

The most notable changes are the rise of polarward meridional velocity near solar equator
and decrease of equatorward velocity at higher latitudes with time. Similar trends can be 
observed within each solar cycle. However it should be noted that the changes are not 
statistically significant due to large errors of the mean velocity near equator and
at higher latitudes due to the smaller number of spots present at these latitudes.
In an attempt to quantify the changes of meridional velocity profiles we calculated linear 
fits trough all datapoints of a given subset. The results of the fit are presented in 
Table \ref{subset}. Slope and intercept with both axes are given. All fits have a
negative slope which is significant over 2$\sigma$ for all solar cycles and all phases of the cycle 
except the minimum and Solar Cycle 22. Also, the intercept with the x-axis (latitude) decreases with
the phase of the cycle, but the change is not statistically significant due to large
errors. A similar result was found by \citet{sudar2014}.

\subsection{Correlation between Meridional Motions and Rotation Rate Residuals
 and Reynolds Stress}

   \begin{figure}
\centering
   \includegraphics[width=0.8\textwidth]{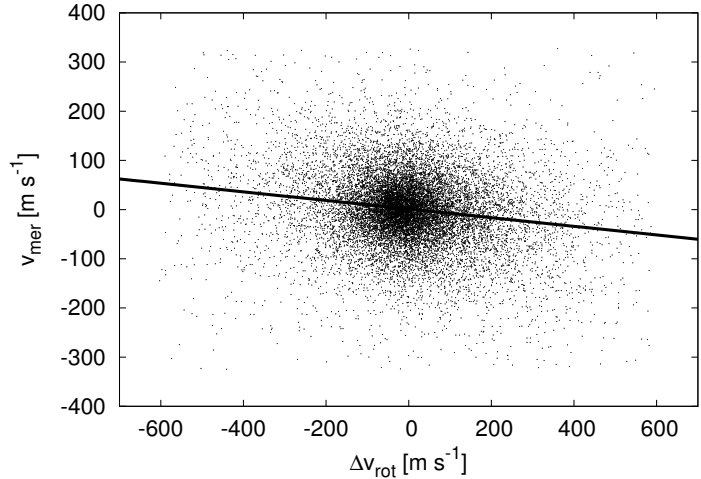}
      \caption{Meridional velocities as a function of rotation rate residuals. Individual data are 
represented with points. The solid line is the linear fit (Equation \ref{linfit}).}
         \label{correlation}
   \end{figure}

To maintain the observed solar differential rotation profile against diffusive
decay, the angular momentum should be somehow transported towards the solar equator. 
This phenomenon can be observed by investigating the relationship between meridional 
velocities and rotation velocity residuals. 
In Figure \ref{correlation} the meridional velocities are plotted against 
the rotation rate residuals. The solid line represents the least square fit
in the form:
\begin{equation}
v_\mathrm{mer} = (-0.0912 \pm 0.0028)\Delta v_\mathrm{rot} + (-0.42 \pm 0.43)\ \mathrm{m\,s}^{-1}.
\label{linfit}
\end{equation}
To check for the influence of outliers on the derived parameters in Equation \ref{linfit} 
the data were also fitted using least deviation method. The least deviation method gives
--0.088 and 0.49 m\,s$^{-1}$ for slope and intercept, respectively.

The slope of the fit is negative indicating that on the average the angular momentum
is transported toward the solar equator. 
   \begin{table}[h]
      \caption{Table of the best fit coefficients (Equation \ref{expfit}).}
         \label{coef}   
         \begin{tabular}{lcr}
            \hline
     Coefficient & Value  & Relative error   \\
                       \hline
  $c_1$ [m$^2$\,s$^{-2}$deg$^{-1}$] & -154$\pm$11 & 7.3\% \\
  $c_3$ [deg$^{-2}$]                & 0.00026$\pm$0.00013 & 50.8\% \\
                       \hline
         \end{tabular}
   \end{table}

   \begin{figure}
\centering
   \includegraphics[width=0.8\textwidth]{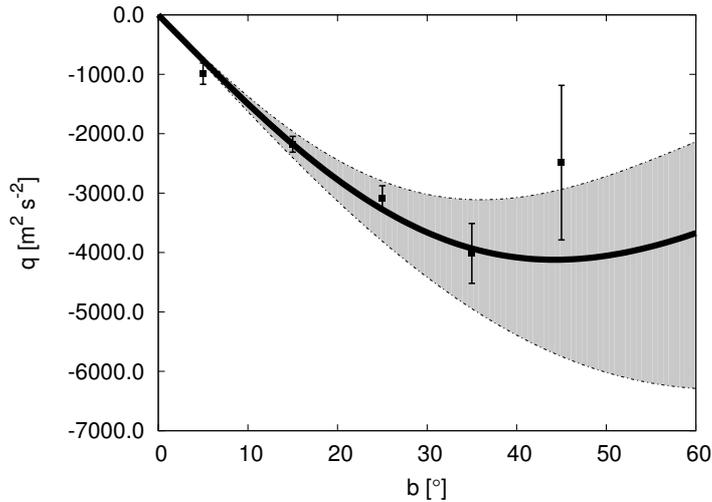}
      \caption{Horizontal component of the Reynolds stress tensor as a function of latitude. 
Data were averaged over 10$^\circ$  in latitude. Solid line represents best fit in the 
form of Equation \ref{expfit}. Shaded areas are defined by errors of the best fit 
coefficients (Table \ref{coef}).
}
         \label{reynolds}
   \end{figure}

The covariance of meridional velocities and rotation velocity residuals gives the horizontal 
component of the Reynolds stress tensor. In Figure \ref{reynolds} the horizontal component
of the Reynolds stress tensor is shown versus latitude. Values were averaged in 10$^\circ$ latitude
stripes. The average values are negative for all latitude stripes which means that the angular
momentum is transported towards lower latitudes, {\it i.e.} toward the solar equator. The solid line in 
Figure \ref{reynolds} represents the empirical exponential cut-off function 
\citep{sudar2014,sudar2017} describing the decreasing trend of the horizontal 
component of the Reynolds stress tensor with latitude:
\begin{equation}
q_{\lambda b}(b)=c_1be^{-c_3b^2},
\label{expfit}
\end{equation}
where $q_{\lambda b}$ is the horizontal component of the Reynolds stress tensor and $b$ is the latitude.
The values of the coefficients $c_1$ and $c_3$ with their respective errors
are given in Table \ref{coef}. The shaded area in the 
figure is defined by the errors of the coefficients $c_1$ and $c_3$.

\section{Discussion and Conclusion}

The differential rotation profile obtained in this work is the same (within 1$\sigma$) 
as the one obtained by \citet{poljancic2017} using the same data and 
(within 2$\sigma$) as by \citet{sudar2017} using the DPD and \citet{sudar2014} using
the GPR and USAF/NOAA datasets. 
The small differences between our result and the result of \citet{poljancic2017}
can be attributed to the different procedures of discarding erroneous values. 
The iterative procedure applied here results in slightly smaller values for rotation 
rate than the 8--19$^\circ$day$^{-1}$ velocity filter used by \citet{poljancic2017}. 
The values of differential
rotation parameters from different methods and datasets are compared in more detail
in \citet{sudar2015} and \citet{poljancic2017}. 

The average values of rotation
rate residuals which do not differ significantly from zero are indicative of the 
quality of the solar rotation profile function fit (Equation \ref{rotprofile}).
Further, our results show meridional motions toward the poles at low latitudes and meridional motions 
toward solar equator at latitudes of 25--30$^\circ$. This is consistent with the picture 
of flows directed toward the centre of activity \citep{sudar2014}. Similar results were 
obtained by \citet{sudar2017} using the DPD dataset.
When each solar hemisphere is treated separately, meridional circulation on southern 
hemisphere is consistent with flows directed toward the centre of activity,
while the flows seem to be predominantly equatorward on the northern hemisphere, 
reminiscent of the flows found by \citet{sivaraman2010} analysing Kodaikanal and Mt. Wilson datasets.
These results confirm that the KSO data 
are of sufficiently high quality and that they can be used in the analysis of solar 
velocity patterns. 

As summarized in \citet{hathaway1996} and \citet{sudar2017} the previously obtained 
results for meridional flows 
are controversial. Both flows toward and away from the centre of activity as well as
flows toward the poles and toward the solar equator were observed. Flows out of
the centre of activity can be attributed to false assumption that any latitude 
(latitude of first, last measurement or mean latitude) can be assigned to the 
observed meridional velocity without taking into account the distribution of tracers. 
This can result in false flows out of the centre of activity \citep{olemskoy2005}. 
Next, the cycle dependence of the meridional motions and rotational velocities 
can influence the results.
More detailed explanation of the above mentioned effects can be found in \citet{sudar2017}.



The difference between flows toward the centre of activity obtained by sunspot 
groups and poleward flows shown by Doppler and CBP data can be reconciled if it 
is assumed that the meridional flow is different in the active regions, where 
sunspots are located, from the flow outside activity areas \citep{sudar2017}.
Alternatively, the anchoring depth of magnetic features can be important, {\it i.e.}
the differences in velocity patterns measured by different features reflect the 
differences in the coupling or anchoring depth of those features. Finally, the
solar meridional flow might be strongly variable \citep{hathaway1996} and the different 
results reflect its variability.

When analysing the meridional motions for possible variations in time and within the
solar cycle motions consistent with flows directed toward the centre of activity were found.
The exception are data for Solar Cycle 22 and the minimum of activity, where the result 
is more reminiscent of equatorward motions. 
The minimum of activity dataset contains the sunspot groups belonging to 
two centers of activity, the one from preceding cycle at low latitudes and the one
from following cycle at higher latitudes, what can influence the result. Besides,
the errors of the meridional velocity values calculated for each latitude stripe
for all subsets (not just minimum) are quite large making most of the values
statistically insignificant and the most notable changes of the profile are at low
latitudes ($<$5$^\circ$) and high latitudes ($>$30$^\circ$) where the number of data
is smallest. Therefore it cannot be concluded that the obtained result represents the actual
changes of meridional motions and is not caused by random error of the mean value for given 
latitude stripe.

By examining the correlation and covariance of meridional velocities and rotation 
rate residuals we found that the angular momentum is transported towards the solar 
equator. The horizontal component of the Reynolds stress tensor is found to be in the 
order of several thousands m$^2$\,s$^{-2}$, with the maximal value of $(-4122\pm 1089)$m$^2$\,s$^{-2}$ 
at $(44\pm11)^\circ$ latitude.
This is in good agreement with the results of other studies using sunspots as tracers
\citep{ward1965,gilman1984,pulkkinen1998b,sudar2014,sudar2017}. This result is also 
in agreement with the theoretical calculations of \citet{canuto1994},
\citet{kapyla2011} and \citet{varela2016}. 
The analysis of the CBP data \citep{vrsnak2003,sudar2016} seems to yield smaller 
values for the horizontal component of the Reynolds stress. As before, this discrepancy 
can be reconciled if it is supposed that the Reynolds stress is stronger around 
active regions. 
This would imply that the major part of angular momentum transfer occurs in
the activity belt. On the other hand, the anchoring depth or height of the tracers
might influence the result, too.

By examining the correlation and covariance of meridional velocities and rotation rate residuals
it was found that the  angular momentum is transported towards the solar equator at all
latitudes. Despite meridional motions and rotation rate residuals having 
values of low statistical significance, their correlation expressed by Reynolds stress
is significant, which means that the Reynolds stress is a robust quantity.
The absolute value of the horizontal component of Reynolds stress is found to be 
increasing from the equator attaining maximum at about 40$^\circ$ latitude which is 
in agreement with results of other studies and theoretical calculations.
The observed values of the Reynolds stress are sufficient to maintain 
the solar differential rotation profile. 
Therefore, our results confirm that the Reynolds stress is the main contributor
to the transport of angular momentum towards solar equator which maintains the
observed solar differential rotation. 
This general result, indicated in various previous studies using other data sets
and methods, is now independently confirmed also by using the KSO data set.
The questions how the anchoring depth of analysed features
and the variability influence the obtained results are still open and
need to be analysed in the future.

\begin{acknowledgements}
This work was partly supported by the Croatian Science Foundation under the project 
6212 ``Solar and Stellar Variability" and in part by the University of Rijeka 
under project number 13.12.1.3.03. We wish to express our  gratitude to 
anonymous referee whose careful review and detailed criticism of the manuscript
helped to improve the presentation and sharpen the arguments.
\end{acknowledgements}

\section*{Disclosure of Potential Conflicts of Interest}

The authors declare that they have no conflicts of interest.

\bibliographystyle{spr-mp-sola}
\bibliography{kanz}

\begin{thebibliography}{41}
\ifx\bisbn     \undefined \def\bisbn  #1{ISBN #1}\fi
\ifx\binits    \undefined \def\binits#1{#1}\fi
\ifx\bauthor   \undefined \def\bauthor#1{#1}\fi
\ifx\batitle   \undefined \def\batitle#1{#1}\fi
\ifx\bjtitle   \undefined \def\bjtitle#1{\textit{#1}}\fi
\ifx\bvolume   \undefined \def\bvolume#1{\textbf{#1}}\fi
\ifx\byear     \undefined \def\byear#1{#1}\fi
\ifx\bissue    \undefined \def\bissue#1{#1}\fi
\ifx\bfpage    \undefined \def\bfpage#1{#1}\fi
\ifx\blpage    \undefined \def\blpage #1{#1}\fi
\ifx\burl      \undefined \def\burl#1{\textsf{#1}}\fi
\ifx\href      \undefined \def\href#1#2{\textsf{#2}}\fi
\ifx\betal     \undefined \def\betal{\textit{et al.}}\fi
\ifx\bctitle   \undefined \def\bctitle#1{#1}\fi
\ifx\beditor   \undefined \def\beditor#1{#1}\fi
\ifx\bbtitle   \undefined \def\bbtitle#1{\textit{#1}}\fi
\ifx\bedition  \undefined \def\bedition#1{#1}\fi
\ifx\bseriesno \undefined \def\bseriesno#1{\textbf{#1}}\fi
\ifx\blocation \undefined \def\blocation#1{#1}\fi
\ifx\bsertitle \undefined \def\bsertitle#1{\textit{#1}}\fi
\ifx\bsnm      \undefined \def\bsnm#1{#1}\fi
\ifx\bsuffix   \undefined \def\bsuffix#1{#1}\fi
\ifx\bparticle \undefined \def\bparticle#1{#1}\fi
\ifx\barticle  \undefined \def\barticle#1{}\fi
\ifx\binstitute  \undefined \def\binstitute#1{#1}\fi
\ifx\bpublisher  \undefined \def\bpublisher#1{#1}\fi
\ifx\doiurl    \undefined
  \def\doiurl#1{\href{http://dx.doi.org/#1}{\textsf{DOI}}}\fi
\ifx\arxivurl  \undefined
  \def\arxivurl#1{\href{http://arxiv.org/abs/#1}{\textsf{arXiv}}}\fi
\ifx\adsurl    \undefined
  \def\adsurl#1{\href{http://adsabs.harvard.edu/abs/#1}{\textsf{ADS}}}\fi
\ifx\botherref \undefined \def\botherref#1{}\fi
\ifx\url       \undefined \def\url#1{\textsf{#1}}\fi
\ifx\bchapter  \undefined \def\bchapter#1{}\fi
\ifx\bbook     \undefined \def\bbook#1{}\fi
\ifx\bcomment  \undefined \def\bcomment#1{#1}\fi
\ifx\oauthor   \undefined \def\oauthor#1{#1}\fi
\ifx\citeauthoryear \undefined\def \citeauthoryear#1{#1}\fi
\ifx\endbibitem\undefined \def\endbibitem{}\fi
\ifx\bconflocation  \undefined \def\bconflocation#1{#1} \fi

\bibitem[\protect\citeauthoryear{{Balthasar} and
  {Fangmeier}}{1988}]{balthasar1988}
\begin{barticle}
\bauthor{\bsnm{{Balthasar}}, \binits{H.}},
\bauthor{\bsnm{{Fangmeier}}, \binits{E.}}:
\byear{1988},
\batitle{{Comparison of the differential rotation laws and meridional motions
  determined from sunspot positions taken from the Greenwich Photoheliographic
  Results, the drawings of G. Spoerer, and the Kanzelhoehe data}}.
\bjtitle{\aap}
\bvolume{203},
\bfpage{381}.
\adsurl{1988A\%26A...203..381B}.
\end{barticle}
\endbibitem

\bibitem[\protect\citeauthoryear{{Balthasar}, {Vazquez}, and
  {W\"ohl}}{1986}]{balthasar1986}
\begin{barticle}
\bauthor{\bsnm{{Balthasar}}, \binits{H.}},
\bauthor{\bsnm{{Vazquez}}, \binits{M.}},
\bauthor{\bsnm{{W\"ohl}}, \binits{H.}}:
\byear{1986},
\batitle{{Differential rotation of sunspot groups in the period from 1874
  through 1976 and changes of the rotation velocity within the solar cycle}}.
\bjtitle{\aap}
\bvolume{155},
\bfpage{87}.
\adsurl{1986A\%26A...155...87B}.
\end{barticle}
\endbibitem

\bibitem[\protect\citeauthoryear{{Braj{\v s}a}
  \textit{et~al.}}{2009}]{brajsa2009}
\begin{barticle}
\bauthor{\bsnm{{Braj{\v s}a}}, \binits{R.}},
\bauthor{\bsnm{{W{\"o}hl}}, \binits{H.}},
\bauthor{\bsnm{{Hanslmeier}}, \binits{A.}},
\bauthor{\bsnm{{Verbanac}}, \binits{G.}},
\bauthor{\bsnm{{Ru{\v z}djak}}, \binits{D.}},
\bauthor{\bsnm{{Cliver}}, \binits{E.}},
\bauthor{\bsnm{{Svalgaard}}, \binits{L.}},
\bauthor{\bsnm{{Roth}}, \binits{M.}}:
\byear{2009},
\batitle{{On solar cycle predictions and reconstructions}}.
\bjtitle{\aap}
\bvolume{496},
\bfpage{855}.
\doiurl{10.1051/0004-6361:200810862}.
\adsurl{2009A\%26A...496..855B}.
\end{barticle}
\endbibitem

\bibitem[\protect\citeauthoryear{{Brun} and {Rempel}}{2009}]{brun2009}
\begin{barticle}
\bauthor{\bsnm{{Brun}}, \binits{A.S.}},
\bauthor{\bsnm{{Rempel}}, \binits{M.}}:
\byear{2009},
\batitle{{Large Scale Flows in the Solar Convection Zone}}.
\bjtitle{\ssr}
\bvolume{144},
\bfpage{151}.
\doiurl{10.1007/s11214-008-9454-9}.
\adsurl{2009SSRv..144..151B}.
\end{barticle}
\endbibitem

\bibitem[\protect\citeauthoryear{{Canuto}, {Minotti}, and
  {Schilling}}{1994}]{canuto1994}
\begin{barticle}
\bauthor{\bsnm{{Canuto}}, \binits{V.M.}},
\bauthor{\bsnm{{Minotti}}, \binits{F.O.}},
\bauthor{\bsnm{{Schilling}}, \binits{O.}}:
\byear{1994},
\batitle{{Differential rotation and turbulent convection: A new Reynolds stress
  model and comparison with solar data}}.
\bjtitle{\apj}
\bvolume{425},
\bfpage{303}.
\doiurl{10.1086/173986}.
\adsurl{1994ApJ...425..303C}.
\end{barticle}
\endbibitem

\bibitem[\protect\citeauthoryear{{Gilman} and {Howard}}{1984}]{gilman1984}
\begin{barticle}
\bauthor{\bsnm{{Gilman}}, \binits{P.A.}},
\bauthor{\bsnm{{Howard}}, \binits{R.}}:
\byear{1984},
\batitle{{On the correlation of longitudinal and latitudinal motions of
  sunspots}}.
\bjtitle{\solphys}
\bvolume{93},
\bfpage{171}.
\doiurl{10.1007/BF00156661}.
\adsurl{1984SoPh...93..171G}.
\end{barticle}
\endbibitem

\bibitem[\protect\citeauthoryear{{Haber} \textit{et~al.}}{2004}]{haber2004}
\begin{barticle}
\bauthor{\bsnm{{Haber}}, \binits{D.A.}},
\bauthor{\bsnm{{Hindman}}, \binits{B.W.}},
\bauthor{\bsnm{{Toomre}}, \binits{J.}},
\bauthor{\bsnm{{Thompson}}, \binits{M.J.}}:
\byear{2004},
\batitle{{Organized Subsurface Flows near Active Regions}}.
\bjtitle{\solphys}
\bvolume{220},
\bfpage{371}.
\doiurl{10.1023/B:SOLA.0000031405.52911.08}.
\adsurl{2004SoPh..220..371H}.
\end{barticle}
\endbibitem

\bibitem[\protect\citeauthoryear{{Hanasoge}
  \textit{et~al.}}{2015}]{hanasoge2015}
\begin{barticle}
\bauthor{\bsnm{{Hanasoge}}, \binits{S.}},
\bauthor{\bsnm{{Miesch}}, \binits{M.S.}},
\bauthor{\bsnm{{Roth}}, \binits{M.}},
\bauthor{\bsnm{{Schou}}, \binits{J.}},
\bauthor{\bsnm{{Sch{\"u}ssler}}, \binits{M.}},
\bauthor{\bsnm{{Thompson}}, \binits{M.J.}}:
\byear{2015},
\batitle{{Solar Dynamics, Rotation, Convection and Overshoot}}.
\bjtitle{\ssr}
\bvolume{196},
\bfpage{79}.
\doiurl{10.1007/s11214-015-0144-0}.
\adsurl{2015SSRv..196...79H}.
\end{barticle}
\endbibitem

\bibitem[\protect\citeauthoryear{{Hanslmeier} and
  {Lustig}}{1986}]{hanslmeier1986}
\begin{barticle}
\bauthor{\bsnm{{Hanslmeier}}, \binits{A.}},
\bauthor{\bsnm{{Lustig}}, \binits{G.}}:
\byear{1986},
\batitle{{Meridional motions of sunspots from 1947.9-1985.0. I - Latitude drift
  at the different solar-cycles}}.
\bjtitle{\aap}
\bvolume{154},
\bfpage{227}.
\adsurl{1986A\%26A...154..227H}.
\end{barticle}
\endbibitem

\bibitem[\protect\citeauthoryear{{Hathaway}}{1996}]{hathaway1996}
\begin{barticle}
\bauthor{\bsnm{{Hathaway}}, \binits{D.H.}}:
\byear{1996},
\batitle{{Doppler Measurements of the Sun's Meridional Flow}}.
\bjtitle{\apj}
\bvolume{460},
\bfpage{1027}.
\doiurl{10.1086/177029}.
\adsurl{1996ApJ...460.1027H}.
\end{barticle}
\endbibitem

\bibitem[\protect\citeauthoryear{{Howard}, {Gilman}, and
  {Gilman}}{1984}]{howard1984}
\begin{barticle}
\bauthor{\bsnm{{Howard}}, \binits{R.}},
\bauthor{\bsnm{{Gilman}}, \binits{P.I.}},
\bauthor{\bsnm{{Gilman}}, \binits{P.A.}}:
\byear{1984},
\batitle{{Rotation of the sun measured from Mount Wilson white-light images}}.
\bjtitle{\apj}
\bvolume{283},
\bfpage{373}.
\doiurl{10.1086/162315}.
\adsurl{1984ApJ...283..373H}.
\end{barticle}
\endbibitem

\bibitem[\protect\citeauthoryear{{Howard}}{1991}]{howard1991}
\begin{barticle}
\bauthor{\bsnm{{Howard}}, \binits{R.F.}}:
\byear{1991},
\batitle{{Cycle latitude effects for sunspot groups}}.
\bjtitle{\solphys}
\bvolume{135},
\bfpage{327}.
\doiurl{10.1007/BF00147504}.
\adsurl{1991SoPh..135..327H}.
\end{barticle}
\endbibitem

\bibitem[\protect\citeauthoryear{{Hr{\v z}ina}
  \textit{et~al.}}{2007}]{sungrabber}
\begin{botherref}
\oauthor{\bsnm{{Hr{\v z}ina}}, \binits{D.}},
\oauthor{\bsnm{{Ro{\v s}a}}, \binits{D.}},
\oauthor{\bsnm{{Hanslmeier}}, \binits{A.}},
\oauthor{\bsnm{{Ru{\v z}djak}}, \binits{V.}},
\oauthor{\bsnm{{Braj{\v s}a}}, \binits{R.}}:
2007,
{Sungrabber - Software for Measurements on Solar Synoptic Images}.
\textit{Central European Astrophysical Bulletin}
\textbf{31}.
\adsurl{2007CEAB...31..273H}.
\end{botherref}
\endbibitem

\bibitem[\protect\citeauthoryear{{K{\"a}pyl{\"a}}
  \textit{et~al.}}{2011}]{kapyla2011}
\begin{barticle}
\bauthor{\bsnm{{K{\"a}pyl{\"a}}}, \binits{P.J.}},
\bauthor{\bsnm{{Mantere}}, \binits{M.J.}},
\bauthor{\bsnm{{Guerrero}}, \binits{G.}},
\bauthor{\bsnm{{Brandenburg}}, \binits{A.}},
\bauthor{\bsnm{{Chatterjee}}, \binits{P.}}:
\byear{2011},
\batitle{{Reynolds stress and heat flux in spherical shell convection}}.
\bjtitle{\aap}
\bvolume{531},
\bfpage{A162}.
\doiurl{10.1051/0004-6361/201015884}.
\adsurl{2011A\%26A...531A.162K}.
\end{barticle}
\endbibitem

\bibitem[\protect\citeauthoryear{{Lustig}}{1983}]{lustig1983}
\begin{barticle}
\bauthor{\bsnm{{Lustig}}, \binits{G.}}:
\byear{1983},
\batitle{{Solar rotation 1947-1981 - Determined from sunspot data}}.
\bjtitle{\aap}
\bvolume{125},
\bfpage{355}.
\adsurl{1983A\%26A...125..355L}.
\end{barticle}
\endbibitem

\bibitem[\protect\citeauthoryear{{Lustig} and {Hanslmeier}}{1987}]{lustig1987}
\begin{barticle}
\bauthor{\bsnm{{Lustig}}, \binits{G.}},
\bauthor{\bsnm{{Hanslmeier}}, \binits{A.}}:
\byear{1987},
\batitle{{Meridional motions of sunspots from 1947.9 to 1985.0. II - Latitude
  motions dependent on SPOT type and phase of the activity cycle}}.
\bjtitle{\aap}
\bvolume{172},
\bfpage{332}.
\adsurl{1987A\%26A...172..332L}.
\end{barticle}
\endbibitem

\bibitem[\protect\citeauthoryear{{Lustig} and {W\"ohl}}{1991}]{lustig1991}
\begin{barticle}
\bauthor{\bsnm{{Lustig}}, \binits{G.}},
\bauthor{\bsnm{{W\"ohl}}, \binits{H.}}:
\byear{1991},
\batitle{{The meridional motions of stable recurrent sunspots}}.
\bjtitle{\aap}
\bvolume{249},
\bfpage{528}.
\adsurl{1991A\%26A...249..528L}.
\end{barticle}
\endbibitem

\bibitem[\protect\citeauthoryear{{Lustig} and {W\"ohl}}{1994}]{lustig1994}
\begin{barticle}
\bauthor{\bsnm{{Lustig}}, \binits{G.}},
\bauthor{\bsnm{{W\"ohl}}, \binits{H.}}:
\byear{1994},
\batitle{{Meridional motions of sunspot groups during eleven activity cycles}}.
\bjtitle{\solphys}
\bvolume{152},
\bfpage{221}.
\doiurl{10.1007/BF01473208}.
\adsurl{1994SoPh..152..221L}.
\end{barticle}
\endbibitem

\bibitem[\protect\citeauthoryear{{Mandal} \textit{et~al.}}{2017}]{mandal2017}
\begin{barticle}
\bauthor{\bsnm{{Mandal}}, \binits{S.}},
\bauthor{\bsnm{{Hegde}}, \binits{M.}},
\bauthor{\bsnm{{Samanta}}, \binits{T.}},
\bauthor{\bsnm{{Hazra}}, \binits{G.}},
\bauthor{\bsnm{{Banerjee}}, \binits{D.}},
\bauthor{\bsnm{{Ravindra}}, \binits{B.}}:
\byear{2017},
\batitle{{Kodaikanal digitized white-light data archive (1921-2011): Analysis
  of various solar cycle features}}.
\bjtitle{\aap}
\bvolume{601},
\bfpage{A106}.
\doiurl{10.1051/0004-6361/201628651}.
\adsurl{2017A\%26A...601A.106M}.
\end{barticle}
\endbibitem

\bibitem[\protect\citeauthoryear{{Olemskoy} and
  {Kitchatinov}}{2005}]{olemskoy2005}
\begin{barticle}
\bauthor{\bsnm{{Olemskoy}}, \binits{S.V.}},
\bauthor{\bsnm{{Kitchatinov}}, \binits{L.L.}}:
\byear{2005},
\batitle{{On the Determination of Meridional Flow on the Sun by the Method of
  Tracers}}.
\bjtitle{Astronomy Letters}
\bvolume{31},
\bfpage{706}.
\doiurl{10.1134/1.2075313}.
\adsurl{2005AstL...31..706O}.
\end{barticle}
\endbibitem

\bibitem[\protect\citeauthoryear{{Poljan\v ci\'c}
  \textit{et~al.}}{2010}]{poljancic2010}
\begin{barticle}
\bauthor{\bsnm{{Poljan\v ci\'c}}, \binits{I.}},
\bauthor{\bsnm{{Braj\v sa}}, \binits{R.}},
\bauthor{\bsnm{{Ru\v zdjak}}, \binits{D.}},
\bauthor{\bsnm{{Hr\v zina}}, \binits{D.}},
\bauthor{\bsnm{{Jurdana-\v Sepi\'c}}, \binits{R.}},
\bauthor{\bsnm{{W\"ohl}}, \binits{H.}},
\bauthor{\bsnm{{Otruba}}, \binits{W.}}:
\byear{2010},
\batitle{{A Comparison of Sunspot Position Measurments from Different Data
  Sets}}.
\bjtitle{Sun and Geosphere}
\bvolume{5},
\bfpage{52}.
\adsurl{2010SunGe...5...52P}.
\end{barticle}
\endbibitem

\bibitem[\protect\citeauthoryear{{Poljan{\v c}i{\'c}}
  \textit{et~al.}}{2011}]{poljancic2011}
\begin{barticle}
\bauthor{\bsnm{{Poljan{\v c}i{\'c}}}, \binits{I.}},
\bauthor{\bsnm{{Braj{\v s}a}}, \binits{R.}},
\bauthor{\bsnm{{Hr{\v z}ina}}, \binits{D.}},
\bauthor{\bsnm{{W{\"o}hl}}, \binits{H.}},
\bauthor{\bsnm{{Hanslmeier}}, \binits{A.}},
\bauthor{\bsnm{{P{\"o}tzi}}, \binits{W.}},
\bauthor{\bsnm{{Baranyi}}, \binits{T.}},
\bauthor{\bsnm{{{\"O}zg{\"u}{\c c}}}, \binits{A.}},
\bauthor{\bsnm{{Singh}}, \binits{J.}},
\bauthor{\bsnm{{Ru{\v z}djak}}, \binits{V.}}:
\byear{2011},
\batitle{{Differences in heliographic positions and rotation velocities of
  sunspot groups from various observatories}}.
\bjtitle{Central European Astrophysical Bulletin}
\bvolume{35},
\bfpage{59}.
\adsurl{2011CEAB...35...59P}.
\end{barticle}
\endbibitem

\bibitem[\protect\citeauthoryear{{Poljan{\v c}i{\'c} Beljan}
  \textit{et~al.}}{2017}]{poljancic2017}
\begin{barticle}
\bauthor{\bsnm{{Poljan{\v c}i{\'c} Beljan}}, \binits{I.}},
\bauthor{\bsnm{{Jurdana-{\v S}epi{\'c}}}, \binits{R.}},
\bauthor{\bsnm{{Braj{\v s}a}}, \binits{R.}},
\bauthor{\bsnm{{Sudar}}, \binits{D.}},
\bauthor{\bsnm{{Ru{\v z}djak}}, \binits{D.}},
\bauthor{\bsnm{{Hr{\v z}ina}}, \binits{D.}},
\bauthor{\bsnm{{P{\"o}tzi}}, \binits{W.}},
\bauthor{\bsnm{{Hanslmeier}}, \binits{A.}},
\bauthor{\bsnm{{Veronig}}, \binits{A.}},
\bauthor{\bsnm{{Skoki{\'c}}}, \binits{I.}},
\bauthor{\bsnm{{W{\"o}hl}}, \binits{H.}}:
\byear{2017},
\batitle{{Solar differential rotation in the period 1964-2016 determined by the
  Kanzelh{\"o}he data set}}.
\bjtitle{\aap}
\bvolume{606},
\bfpage{A72}.
\doiurl{10.1051/0004-6361/201731047}.
\adsurl{2017A\%26A...606A..72P}.
\end{barticle}
\endbibitem

\bibitem[\protect\citeauthoryear{{Pulkkinen} and
  {Tuominen}}{1998a}]{pulkkinen1998a}
\begin{barticle}
\bauthor{\bsnm{{Pulkkinen}}, \binits{P.}},
\bauthor{\bsnm{{Tuominen}}, \binits{I.}}:
\byear{1998}a,
\batitle{{Velocity structures from sunspot statistics in cycles 10 to 22. I.
  Rotational velocity}}.
\bjtitle{\aap}
\bvolume{332},
\bfpage{748}.
\adsurl{1998A\%26A...332..748P}.
\end{barticle}
\endbibitem

\bibitem[\protect\citeauthoryear{{Pulkkinen} and
  {Tuominen}}{1998b}]{pulkkinen1998b}
\begin{barticle}
\bauthor{\bsnm{{Pulkkinen}}, \binits{P.}},
\bauthor{\bsnm{{Tuominen}}, \binits{I.}}:
\byear{1998}b,
\batitle{{Velocity structures from sunspot statistics in cycles 10 to 22. II.
  Latitudinal velocity and correlation functions}}.
\bjtitle{\aap}
\bvolume{332},
\bfpage{755}.
\adsurl{1998A\%26A...332..755P}.
\end{barticle}
\endbibitem

\bibitem[\protect\citeauthoryear{{R{\"u}diger} and
  {Hollerbach}}{2004}]{Rudiger2004}
\begin{bbook}
\bauthor{\bsnm{{R{\"u}diger}}, \binits{G.}},
\bauthor{\bsnm{{Hollerbach}}, \binits{R.}}:
\byear{2004},
\bbtitle{{The magnetic universe : geophysical and astrophysical dynamo
  theory}},
\bfpage{343}.
\adsurl{2004muga.book.....R}.
\end{bbook}
\endbibitem

\bibitem[\protect\citeauthoryear{{Ru{\v z}djak}
  \textit{et~al.}}{2004}]{ruzdjak2004}
\begin{barticle}
\bauthor{\bsnm{{Ru{\v z}djak}}, \binits{D.}},
\bauthor{\bsnm{{Ru{\v z}djak}}, \binits{V.}},
\bauthor{\bsnm{{Braj{\v s}a}}, \binits{R.}},
\bauthor{\bsnm{{W{\"o}hl}}, \binits{H.}}:
\byear{2004},
\batitle{{Deceleration of the rotational velocities of sunspot groups during
  their evolution}}.
\bjtitle{\solphys}
\bvolume{221},
\bfpage{225}.
\doiurl{10.1023/B:SOLA.0000035066.96031.4f}.
\adsurl{2004SoPh..221..225R}.
\end{barticle}
\endbibitem

\bibitem[\protect\citeauthoryear{{Sivaraman}
  \textit{et~al.}}{2010}]{sivaraman2010}
\begin{barticle}
\bauthor{\bsnm{{Sivaraman}}, \binits{K.R.}},
\bauthor{\bsnm{{Sivaraman}}, \binits{H.}},
\bauthor{\bsnm{{Gupta}}, \binits{S.S.}},
\bauthor{\bsnm{{Howard}}, \binits{R.F.}}:
\byear{2010},
\batitle{{Return Meridional Flow in the Convection Zone from Latitudinal
  Motions of Umbrae of Sunspot Groups}}.
\bjtitle{\solphys}
\bvolume{266},
\bfpage{247}.
\doiurl{10.1007/s11207-010-9620-6}.
\adsurl{2010SoPh..266..247S}.
\end{barticle}
\endbibitem

\bibitem[\protect\citeauthoryear{{Skoki{\'c}}
  \textit{et~al.}}{2014}]{skokic2014}
\begin{barticle}
\bauthor{\bsnm{{Skoki{\'c}}}, \binits{I.}},
\bauthor{\bsnm{{Braj{\v s}a}}, \binits{R.}},
\bauthor{\bsnm{{Ro{\v s}a}}, \binits{D.}},
\bauthor{\bsnm{{Hr{\v z}ina}}, \binits{D.}},
\bauthor{\bsnm{{W{\"o}hl}}, \binits{H.}}:
\byear{2014},
\batitle{{Validity of the Relations Between the Synodic and Sidereal Rotation
  Velocities of the Sun}}.
\bjtitle{\solphys}
\bvolume{289},
\bfpage{1471}.
\doiurl{10.1007/s11207-013-0426-1}.
\adsurl{2014SoPh..289.1471S}.
\end{barticle}
\endbibitem

\bibitem[\protect\citeauthoryear{{Sudar} \textit{et~al.}}{2014}]{sudar2014}
\begin{barticle}
\bauthor{\bsnm{{Sudar}}, \binits{D.}},
\bauthor{\bsnm{{Skoki{\'c}}}, \binits{I.}},
\bauthor{\bsnm{{Ru{\v z}djak}}, \binits{D.}},
\bauthor{\bsnm{{Braj{\v s}a}}, \binits{R.}},
\bauthor{\bsnm{{W{\"o}hl}}, \binits{H.}}:
\byear{2014},
\batitle{{Tracing sunspot groups to determine angular momentum transfer on the
  Sun}}.
\bjtitle{\mnras}
\bvolume{439},
\bfpage{2377}.
\doiurl{10.1093/mnras/stu099}.
\adsurl{2014MNRAS.439.2377S}.
\end{barticle}
\endbibitem

\bibitem[\protect\citeauthoryear{{Sudar} \textit{et~al.}}{2015}]{sudar2015}
\begin{barticle}
\bauthor{\bsnm{{Sudar}}, \binits{D.}},
\bauthor{\bsnm{{Skoki{\'c}}}, \binits{I.}},
\bauthor{\bsnm{{Braj{\v s}a}}, \binits{R.}},
\bauthor{\bsnm{{Saar}}, \binits{S.H.}}:
\byear{2015},
\batitle{{Steps towards a high precision solar rotation profile: Results from
  SDO/AIA coronal bright point data}}.
\bjtitle{\aap}
\bvolume{575},
\bfpage{A63}.
\doiurl{10.1051/0004-6361/201424929}.
\adsurl{2015A\%26A...575A..63S}.
\end{barticle}
\endbibitem

\bibitem[\protect\citeauthoryear{{Sudar} \textit{et~al.}}{2016}]{sudar2016}
\begin{barticle}
\bauthor{\bsnm{{Sudar}}, \binits{D.}},
\bauthor{\bsnm{{Saar}}, \binits{S.H.}},
\bauthor{\bsnm{{Skoki{\'c}}}, \binits{I.}},
\bauthor{\bsnm{{Poljan{\v c}i{\'c} Beljan}}, \binits{I.}},
\bauthor{\bsnm{{Braj{\v s}a}}, \binits{R.}}:
\byear{2016},
\batitle{{Meridional motions and Reynolds stress from SDO/AIA coronal bright
  points data}}.
\bjtitle{\aap}
\bvolume{587},
\bfpage{A29}.
\doiurl{10.1051/0004-6361/201527217}.
\adsurl{2016A\%26A...587A..29S}.
\end{barticle}
\endbibitem

\bibitem[\protect\citeauthoryear{{Sudar} \textit{et~al.}}{2017}]{sudar2017}
\begin{barticle}
\bauthor{\bsnm{{Sudar}}, \binits{D.}},
\bauthor{\bsnm{{Braj{\v s}a}}, \binits{R.}},
\bauthor{\bsnm{{Skoki{\'c}}}, \binits{I.}},
\bauthor{\bsnm{{Poljan{\v c}i{\'c} Beljan}}, \binits{I.}},
\bauthor{\bsnm{{W{\"o}hl}}, \binits{H.}}:
\byear{2017},
\batitle{{Meridional Motion and Reynolds Stress from Debrecen Photoheliographic
  Data}}.
\bjtitle{\solphys}
\bvolume{292},
\bfpage{86}.
\doiurl{10.1007/s11207-017-1105-4}.
\adsurl{2017SoPh..292...86S}.
\end{barticle}
\endbibitem

\bibitem[\protect\citeauthoryear{{{\v S}vanda}, {Kosovichev}, and
  {Zhao}}{2008}]{svanda2008}
\begin{barticle}
\bauthor{\bsnm{{{\v S}vanda}}, \binits{M.}},
\bauthor{\bsnm{{Kosovichev}}, \binits{A.G.}},
\bauthor{\bsnm{{Zhao}}, \binits{J.}}:
\byear{2008},
\batitle{{Effects of Solar Active Regions on Meridional Flows}}.
\bjtitle{\apjl}
\bvolume{680},
\bfpage{L161}.
\doiurl{10.1086/589997}.
\adsurl{2008ApJ...680L.161S}.
\end{barticle}
\endbibitem

\bibitem[\protect\citeauthoryear{{Varela}, {Strugarek}, and
  {Brun}}{2016}]{varela2016}
\begin{barticle}
\bauthor{\bsnm{{Varela}}, \binits{J.}},
\bauthor{\bsnm{{Strugarek}}, \binits{A.}},
\bauthor{\bsnm{{Brun}}, \binits{A.S.}}:
\byear{2016},
\batitle{{Characterizing the feedback of magnetic field on the differential
  rotation of solar-like stars}}.
\bjtitle{Advances in Space Research}
\bvolume{58},
\bfpage{1507}.
\doiurl{10.1016/j.asr.2016.06.032}.
\adsurl{2016AdSpR..58.1507V}.
\end{barticle}
\endbibitem

\bibitem[\protect\citeauthoryear{{Veronig} and {P{\"o}tzi}}{2016}]{veronig2016}
\begin{bchapter}
\bauthor{\bsnm{{Veronig}}, \binits{A.M.}},
\bauthor{\bsnm{{P{\"o}tzi}}, \binits{W.}}:
\byear{2016},
\bctitle{{Ground-based Observations of the Solar Sources of Space Weather}}.
In: \beditor{\bsnm{{Dorotovic}}, \binits{I.}},
\beditor{\bsnm{{Fischer}}, \binits{C.E.}},
\beditor{\bsnm{{Temmer}}, \binits{M.}} (eds.)
\bbtitle{Coimbra Solar Physics Meeting: Ground-based Solar Observations in the
  Space Instrumentation Era},
\bsertitle{Astronomical Society of the Pacific Conference Series}
\bseriesno{504},
\bfpage{247}.
\adsurl{2016ASPC..504..247V}.
\end{bchapter}
\endbibitem

\bibitem[\protect\citeauthoryear{{Vr{\v s}nak}
  \textit{et~al.}}{2003}]{vrsnak2003}
\begin{barticle}
\bauthor{\bsnm{{Vr{\v s}nak}}, \binits{B.}},
\bauthor{\bsnm{{Braj{\v s}a}}, \binits{R.}},
\bauthor{\bsnm{{W{\"o}hl}}, \binits{H.}},
\bauthor{\bsnm{{Ru{\v z}djak}}, \binits{V.}},
\bauthor{\bsnm{{Clette}}, \binits{F.}},
\bauthor{\bsnm{{Hochedez}}, \binits{J.-F.}}:
\byear{2003},
\batitle{{Properties of the solar velocity field indicated by motions of
  coronal bright points}}.
\bjtitle{\aap}
\bvolume{404},
\bfpage{1117}.
\doiurl{10.1051/0004-6361:20030502}.
\adsurl{2003A\%26A...404.1117V}.
\end{barticle}
\endbibitem

\bibitem[\protect\citeauthoryear{{Ward}}{1965}]{ward1965}
\begin{barticle}
\bauthor{\bsnm{{Ward}}, \binits{F.}}:
\byear{1965},
\batitle{{The General Circulation of the Solar Atmosphere and the Maintenance
  of the Equatorial Acceleration.}}
\bjtitle{\apj}
\bvolume{141},
\bfpage{534}.
\doiurl{10.1086/148143}.
\adsurl{1965ApJ...141..534W}.
\end{barticle}
\endbibitem

\bibitem[\protect\citeauthoryear{{Watson} and {Fletcher}}{2011}]{watson2011}
\begin{bchapter}
\bauthor{\bsnm{{Watson}}, \binits{F.}},
\bauthor{\bsnm{{Fletcher}}, \binits{L.}}:
\byear{2011},
\bctitle{{Automated sunspot detection and the evolution of sunspot magnetic
  fields during solar cycle 23}}.
In: \beditor{\bsnm{{Prasad Choudhary}}, \binits{D.}},
\beditor{\bsnm{{Strassmeier}}, \binits{K.G.}} (eds.)
\bbtitle{Physics of Sun and Star Spots},
\bsertitle{IAU Symposium}
\bseriesno{273},
\bfpage{51}.
\doiurl{10.1017/S1743921311014992}.
\adsurl{2011IAUS..273...51W}.
\end{bchapter}
\endbibitem

\bibitem[\protect\citeauthoryear{{W{\"o}hl} and {Braj{\v
  s}a}}{2001}]{woehl2001}
\begin{barticle}
\bauthor{\bsnm{{W{\"o}hl}}, \binits{H.}},
\bauthor{\bsnm{{Braj{\v s}a}}, \binits{R.}}:
\byear{2001},
\batitle{{Meridional Motions of Stable Recurrent Sunspot Groups}}.
\bjtitle{\solphys}
\bvolume{198},
\bfpage{57}.
\doiurl{10.1023/A:1005244301820}.
\adsurl{2001SoPh..198...57W}.
\end{barticle}
\endbibitem

\bibitem[\protect\citeauthoryear{{Zuccarello} and
  {Zappal{\'a}}}{2003}]{zuccarello2003}
\begin{barticle}
\bauthor{\bsnm{{Zuccarello}}, \binits{F.}},
\bauthor{\bsnm{{Zappal{\'a}}}, \binits{R.A.}}:
\byear{2003},
\batitle{{Angular velocity during the cycle deduced using the sunspot group age
  selection methodology}}.
\bjtitle{Astronomische Nachrichten}
\bvolume{324},
\bfpage{464}.
\doiurl{10.1002/asna.200310155}.
\adsurl{2003AN....324..464Z}.
\end{barticle}
\endbibitem

\end{thebibliography}

\end{article} 

\end{document}